\documentclass[12pt]{article}
\pdfoutput=1
\usepackage{a4wide,epsfig,psfrag,amsmath,amssymb,cite,scalefnt}
\usepackage{color}
\usepackage{amsmath,comment,braket}
\usepackage{placeins}

\graphicspath{ {figs/} }

\newcommand{\be}{\begin{equation}}
\newcommand{\ee}{\end{equation}}
\newcommand{\bea}{\begin{eqnarray}}
\newcommand{\eea}{\end{eqnarray}}

\allowdisplaybreaks

\newcommand{\gsim}{\;\rlap{\lower 3.5 pt \hbox{$\mathchar \sim$}} \raise 1pt
 \hbox {$>$}\;}
\newcommand{\lsim}{\;\rlap{\lower 3.5 pt \hbox{$\mathchar \sim$}} \raise 1pt
 \hbox {$<$}\;}

\begin{document}

\title{\vskip-3cm{\baselineskip14pt
    \begin{flushleft}
      \normalsize 
      TTP20-006\\
      P3H-20-007\\
  \end{flushleft}}
  \vskip1.5cm
  $gg\to ZZ$: analytic two-loop results for the low- and high-energy regions
}

\author{
  Joshua Davies$^{a}$,
  Go Mishima$^{a,b}$,
  Matthias Steinhauser$^{a}$,
  David Wellmann$^{a}$
  \\[1mm]
  {\small\it $^a$Institut f{\"u}r Theoretische Teilchenphysik}\\
  {\small\it Karlsruhe Institute of Technology (KIT)}\\
  {\small\it Wolfgang-Gaede Stra\ss{}e 1, 76128 Karlsruhe, Germany}
  \\[1mm]
  {\small\it $^b$Institut f{\"u}r Kernphysik}\\
  {\small\it Karlsruhe Institute of Technology (KIT)}
  \\
  {\small\it Hermann-von-Helmholtz-Platz 1, 76344 Eggenstein-Leopoldshafen, Germany}
}

\date{}

\maketitle

\thispagestyle{empty}

\begin{abstract}

  We compute next-to-leading order virtual two-loop corrections to the process
  $gg\to ZZ$ in the low- and high-energy limits, considering the contributions
  with virtual top quarks. Analytic results for all 20 form factors are
  presented including expansion terms up to $1/m_t^{12}$ and
  $m_t^{32}$. We use a Pad\'e approximation procedure to extend the radius of
  convergence of the high-energy expansion and apply this approach to the finite virtual
  next-to-leading order corrections.

% \medskip
%
% \noindent
% PACS numbers:

\end{abstract}

\thispagestyle{empty}

\sloppy

%- }}}

\newpage

%- {{{ Introduction:

\section{Introduction}

The production of $Z$ boson pairs constitutes an important process at the CERN
Large Hadron Collider (LHC). It can be measured with an accuracy of a few
percent (see, e.g., Ref.~\cite{Aaboud:2019lgy,Sirunyan:2017zjc}) and,
furthermore, plays an important role both for on-shell and off-shell Higgs
boson production. The latter is particularly important in the context of the
indirect determination of the Higgs boson
width~\cite{Aaboud:2018puo,Sirunyan:2019twz}, as was pointed out in
Refs.~\cite{Kauer:2012hd,Caola:2013yja,Campbell:2013una}.

In recent years there has been quite some activity on the theory side with
the aim to compute higher order corrections which enable precise predictions.
At tree level $ZZ$ production proceeds via quark--anti-quark annihilation
where NNLO corrections are
available~\cite{Cascioli:2014yka,Heinrich:2017bvg,Gehrmann:2014bfa,Caola:2014iua,Gehrmann:2015ora,Grazzini:2015hta,Kallweit:2018nyv}.

The gluon fusion channel is loop induced and is thus formally of NNLO. It
turns out that the one-loop contribution~\cite{Glover:1988rg} from massless quarks is quite
large and amounts to more than half of the NNLO
contribution~\cite{Cascioli:2014yka}.  NLO (two-loop) QCD corrections to
$gg\to ZZ$ with massless quarks have been computed in
Refs.~\cite{vonManteuffel:2015msa,Caola:2015psa}.  A large $K$-factor of
50\nobreakdash-100\% (depending of the renormalization and factorization scales) has been
observed which increases the $pp\to ZZ$ cross section by about
5\%~\cite{Grazzini:2018owa}.

The top quark contribution to $gg\to ZZ$ is expected to be particularly
relevant for higher invariant masses providing a relevant impact on the
indirect determination of the Higgs boson
width~\cite{Kauer:2012hd,Caola:2013yja,Campbell:2013una}.  Its computation is
technically more challenging than the massless counterpart and currently only
the one-loop corrections are available in exact form~\cite{Glover:1988rg}.
Exact two-loop corrections with virtual top quarks are not yet available,
however, approximations have been considered by several groups.  The leading
term in the large top quark mass expansion has been considered
in~\cite{Melnikov:2015laa}.
In Ref.~\cite{Campbell:2016ivq} the interference of $gg \to ZZ$
with $gg\to H \to ZZ$ has been computed, in an expansion up to  $1/m_t^{12}$.
A conformal mapping and Pad\'e
approximation have been applied with the aim to extend the validity of the
large-mass expansion. Furthermore, the (anomalous) double triangle
contributions have been computed with exact dependence on the masses and
kinematic variables.  Recently, in Ref.~\cite{Grober:2019kuf} conformal
mapping and Pad\'e approximation have been used in order to combine information
from the large-$m_t$ and the threshold regions. Also in this work results are
presented for the interference to the off-shell Higgs contributions.

In this work we concentrate on the loop-induced gluon fusion channel with
virtual top quarks.  Its leading (one-loop) term is already a NNLO contribution
to $pp\to ZZ$. It amounts to a few percent of the  numerically large massless
contribution and it is thus desirable to compute the two-loop terms, which formally
are N$^3$LO.

The contributing Feynman diagrams (see Fig.~\ref{fig::sample} for a few
examples) can be subdivided into triangle and box contributions, where the
former corresponds to $gg\to H \to ZZ$, i.e., a virtual Higgs boson connects
the quark loop and the final-state $Z$ boson. Exact results for the
Higgs-gluon vertex corrections up to two loops are known
from~\cite{Harlander:2005rq,Anastasiou:2006hc,Aglietti:2006tp}. 

In this paper we compute analytic one- and two-loop results of the top quark
contribution for all 20 form factors. We choose an orthogonal basis which
simplifies the computation of the squared amplitude.
Expressing the final result as a linear combination of form factors provides
full flexibility; for example, it is straightforward to compute the projection
on the Higgs-induced sub-process $gg\to H\to ZZ$. In an alternative approach
we also express our results in terms of helicity amplitudes (see, e.g.,
Ref.~\cite{vonManteuffel:2015msa,Dixon:2013uaa}).
We consider an expansion for
both large and small top quark masses. In the latter case we take finite $Z$
boson masses into account by a subsequent expansion in $m_Z^2/m_t^2$.  Parts
of our large-$m_t$ results can be compared to
Refs.~\cite{Melnikov:2015laa,Campbell:2016ivq} whereas the high-energy results
are new.

We do not consider the two-loop light-quark contributions, which are known
from~\cite{vonManteuffel:2015msa,Caola:2015ila}.
Similarly, we do not consider the contribution originating from two
quark triangles, which has been computed in~\cite{Campbell:2016ivq}.  We also
do not compute real radiation contributions in this paper, but concentrate
on the virtual corrections.

The remainder of the paper is organized as follows. In the next section we
introduce our definitions and notation, and describe our methodology for
the computation of the high-energy and large-$m_t$ expansions. We also
discuss how one can obtain helicity amplitudes from our form factors.
In Section~\ref{sec::LO} we compare the expansions to the exact LO result
and justify our choices for the expansion depths used at NLO. In Section~
\ref{sec::pade} we describe how one can improve the radius of convergence
of the high-energy expansions by making use of Pad\'e approximants.
Using this method, in Section~\ref{sec::NLO} we show NLO results for
form factors and for the finite virtual corrections to the cross section.
For the latter, we consider different values for the transverse momentum of
the $Z$ bosons and demonstrate that we can obtain stable predictions for
this quantity for transverse momentum values as small as $150$~GeV.
Our conclusions are presented in Section~\ref{sec::concl}.
In the Appendix we provide the explicit results for the relations which can be
used to rotate to the orthogonal tensor basis of Section~\ref{sec::orthog-tensor-basis}.
Furthermore, numerical results for all LO and NLO form factors and analytic results for
some example LO form factors are presented.

%- }}}
%- {{{ Technicalities and high-energy expansion:

\section{\label{sec::technical}Technical details}

\begin{figure}[t]
\begin{center}
  \begin{tabular}{ccc}
  \includegraphics[width=0.22\textwidth]{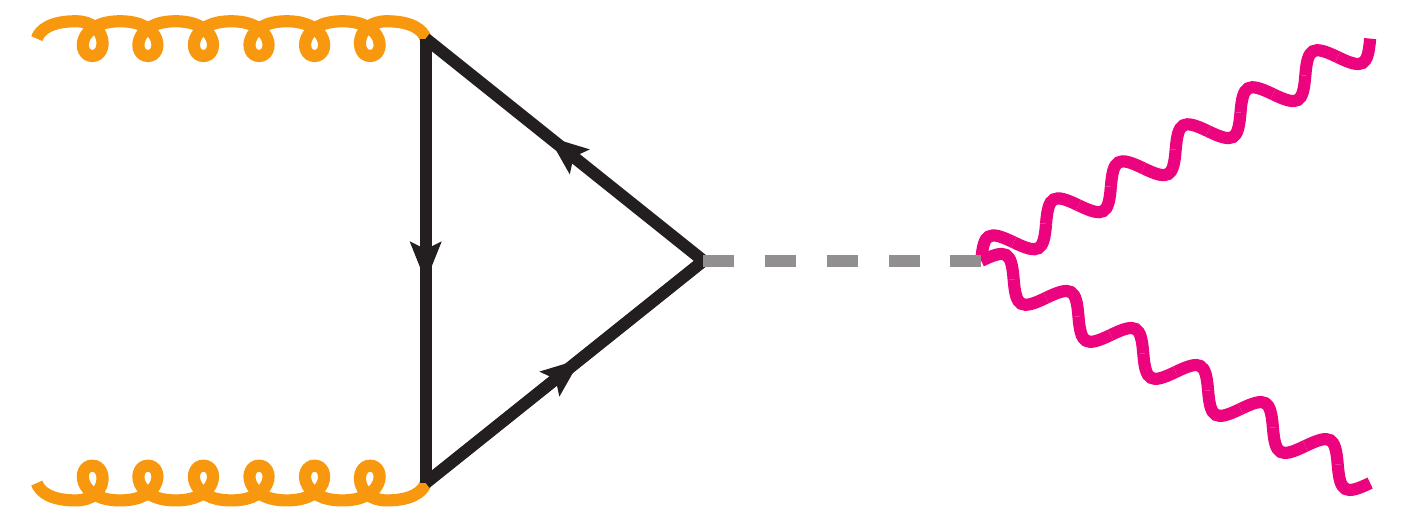} &
  \includegraphics[width=0.22\textwidth]{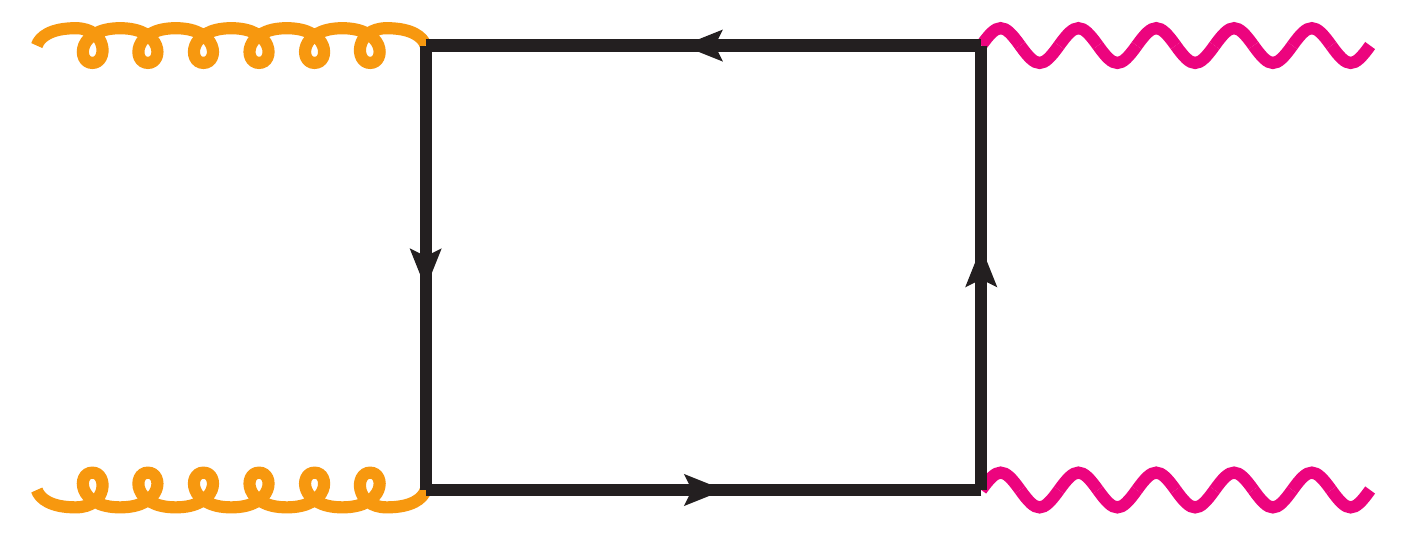} &
  \includegraphics[width=0.22\textwidth]{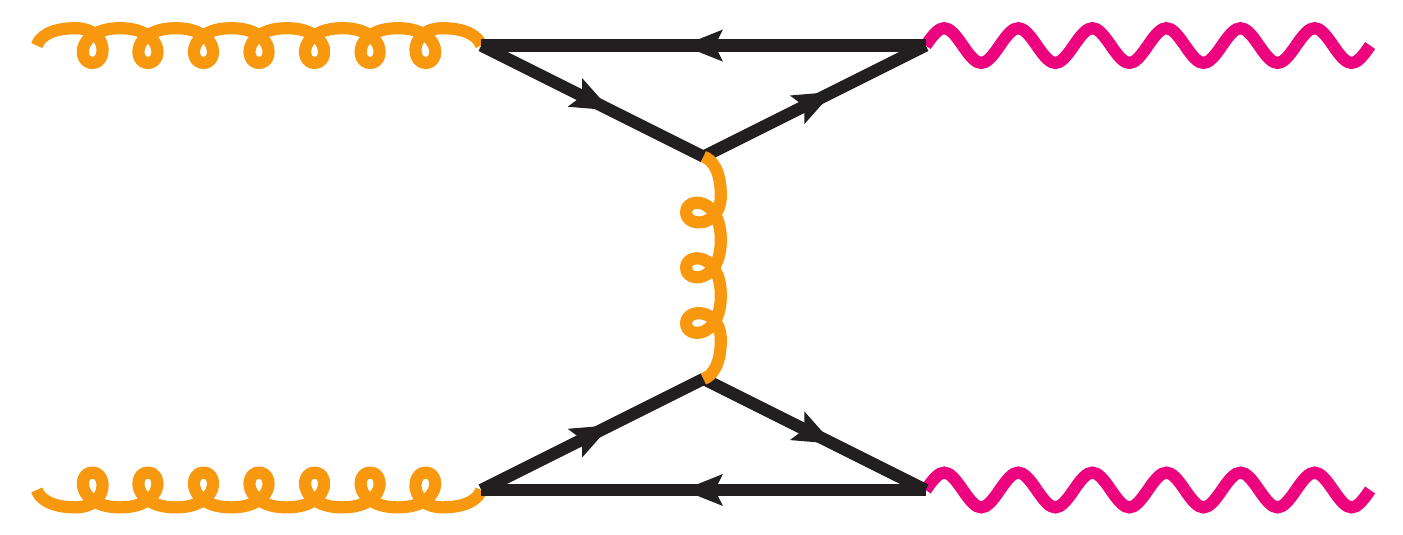} \\
  \includegraphics[width=0.22\textwidth]{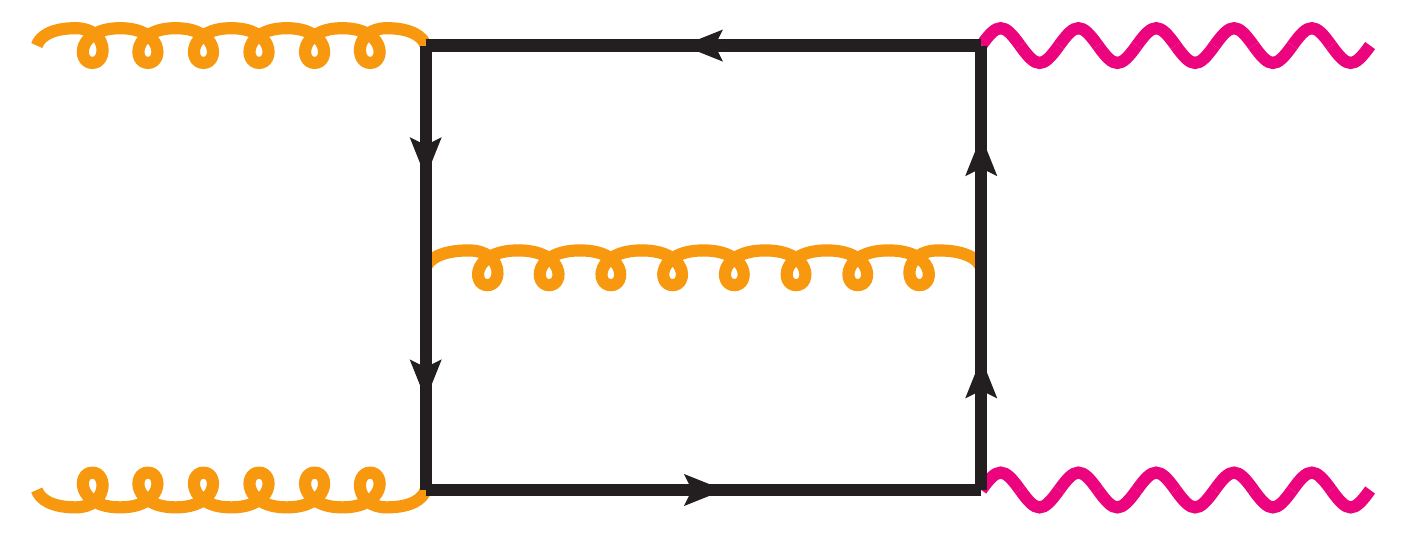} &
  \includegraphics[width=0.22\textwidth]{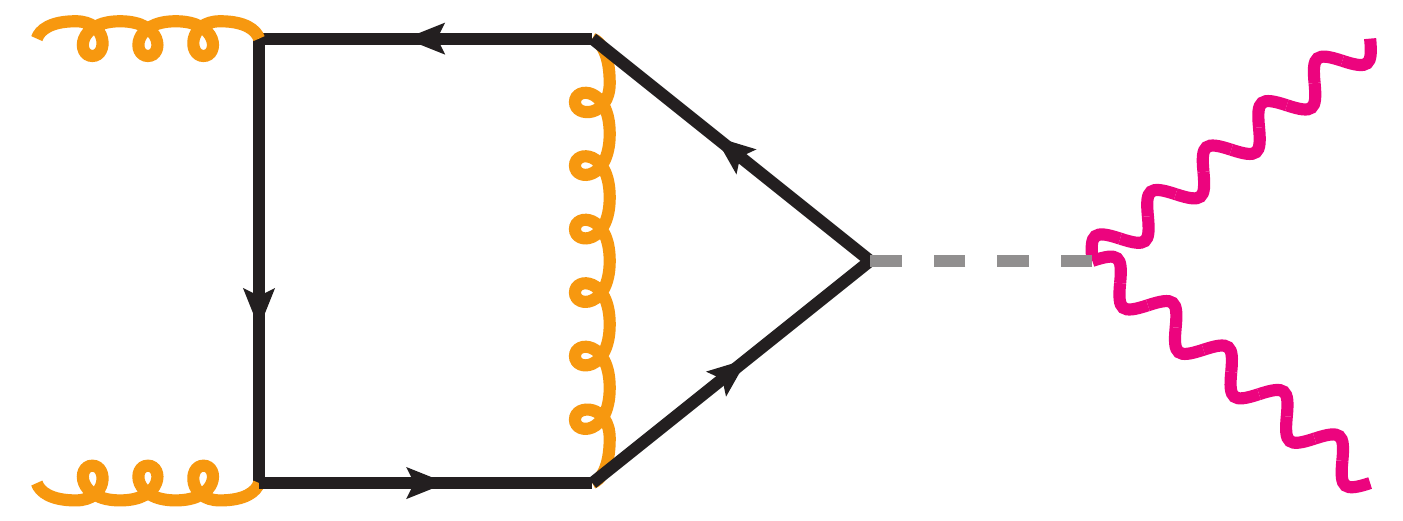} &
  \includegraphics[width=0.22\textwidth]{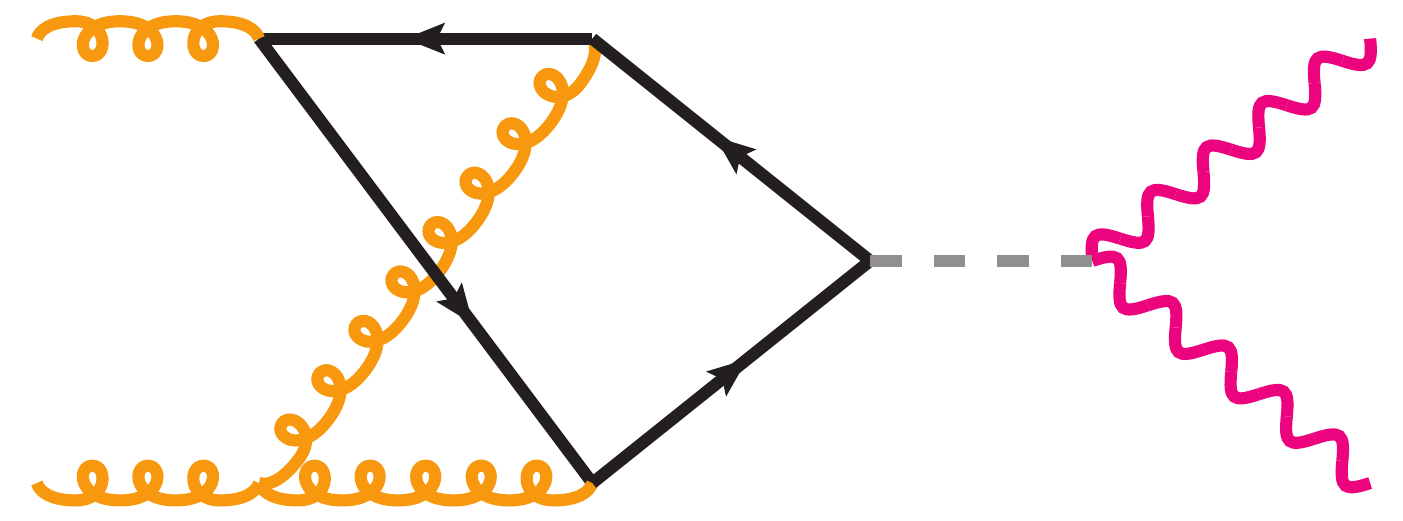} \\
  \includegraphics[width=0.22\textwidth]{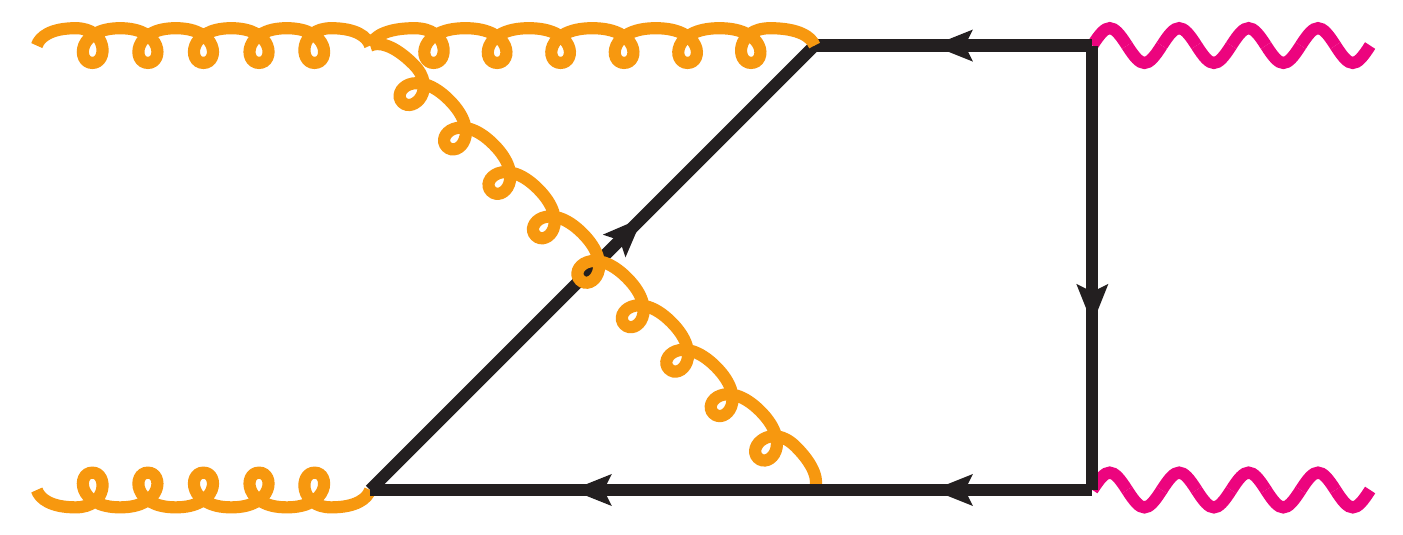} &
  \includegraphics[width=0.22\textwidth]{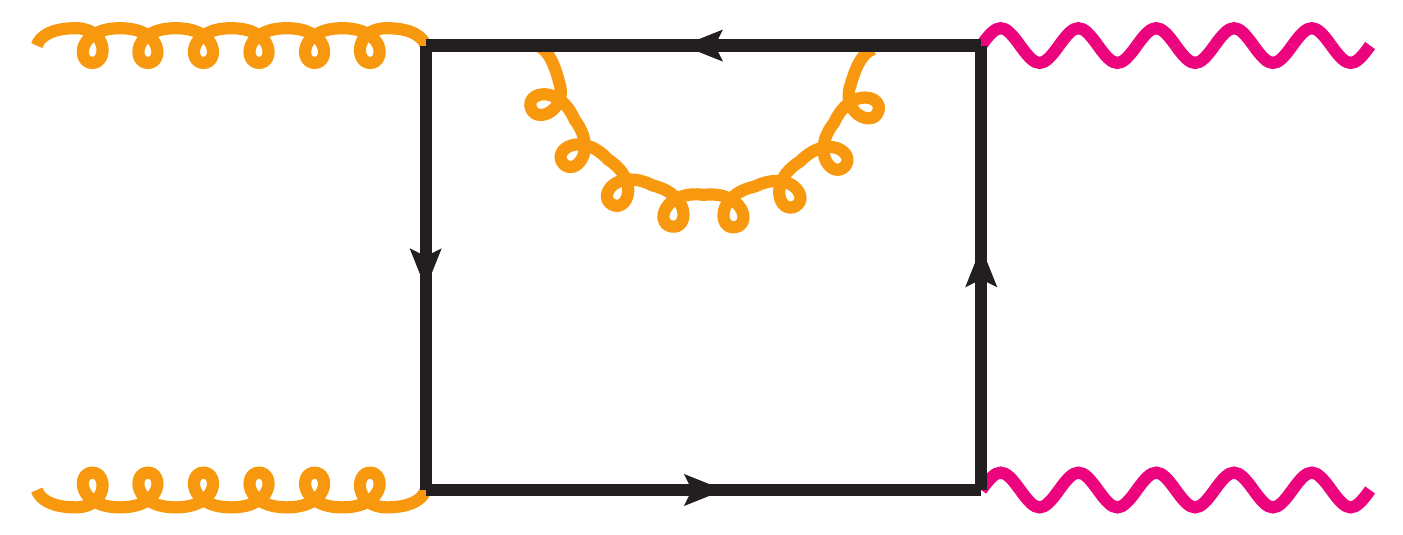} &
  \includegraphics[width=0.22\textwidth]{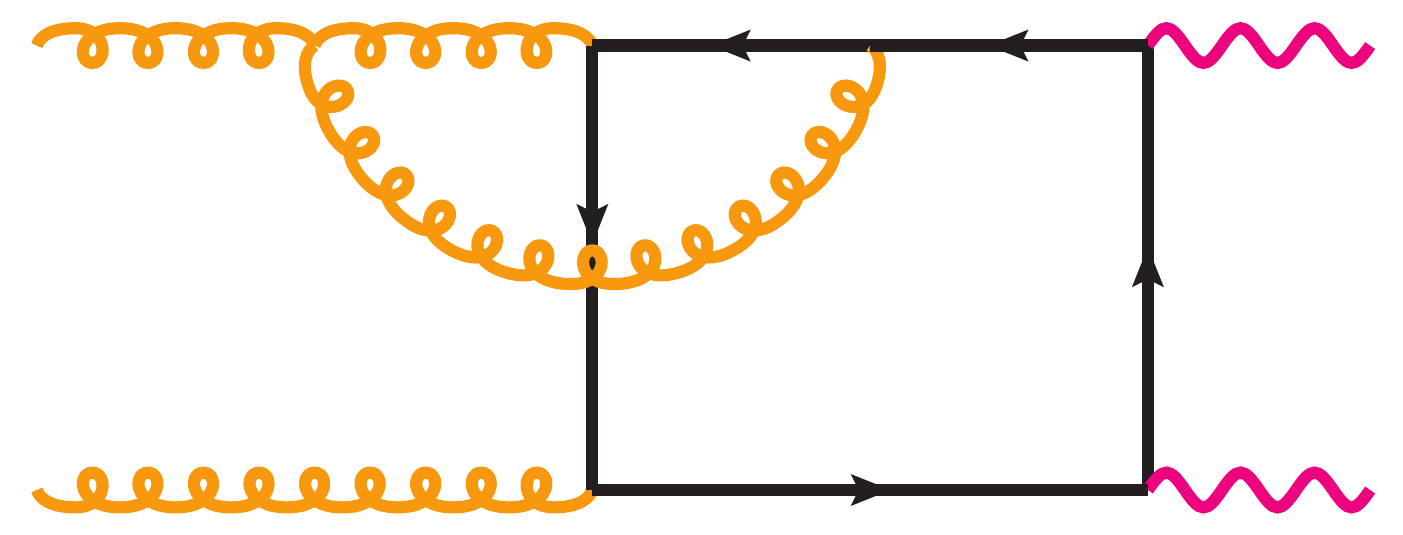} \\
  \includegraphics[width=0.22\textwidth]{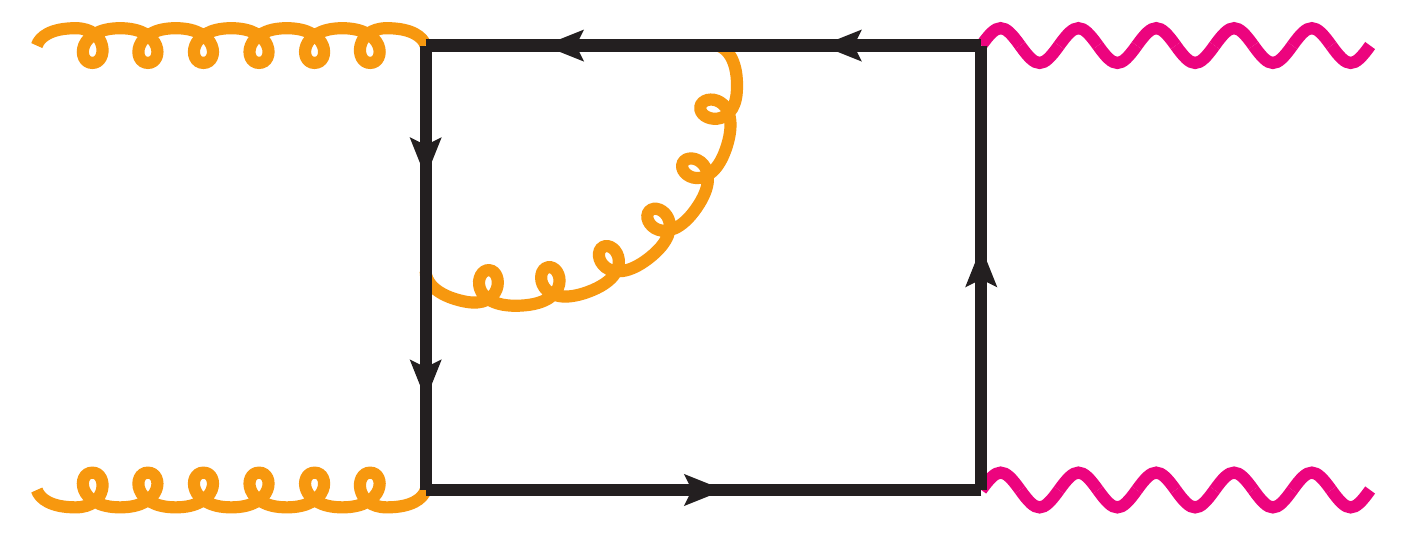} &
  \includegraphics[width=0.22\textwidth]{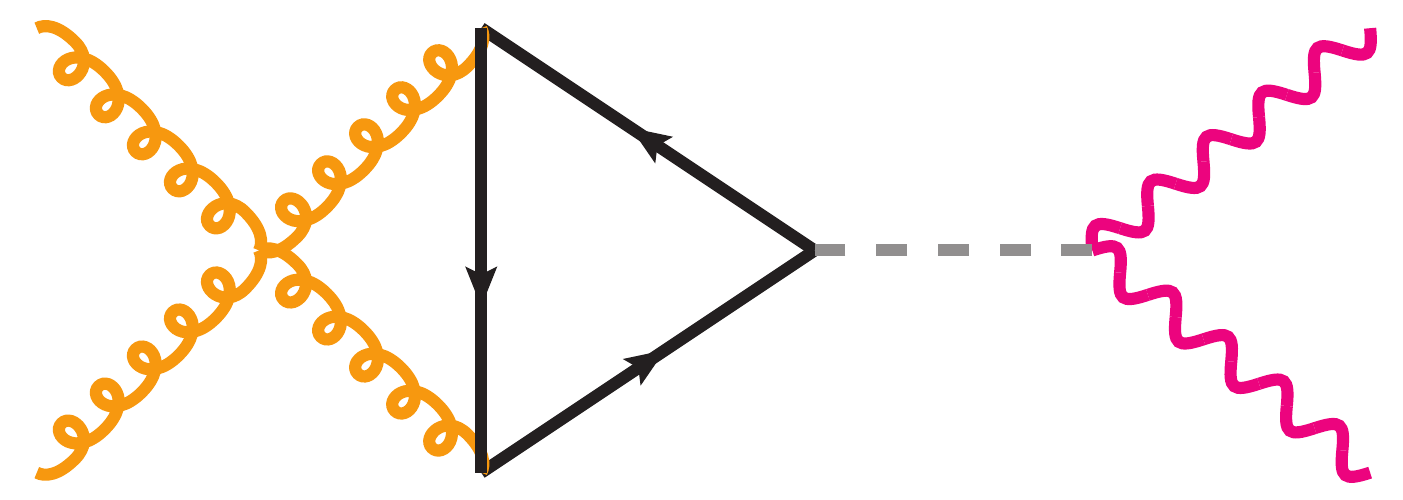} &
  \includegraphics[width=0.22\textwidth]{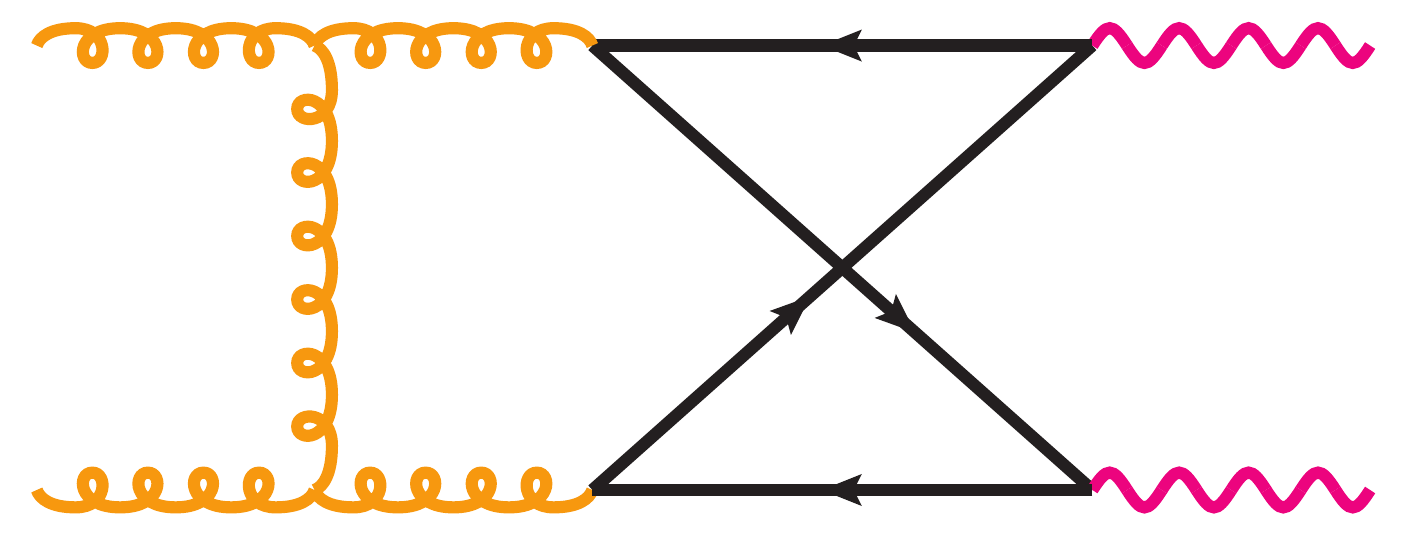} \\
  \end{tabular}
  \caption{\label{fig::sample}Sample LO and NLO Feynman diagrams for
    $gg\to ZZ$. ``Double triangle'' diagrams (such as the third diagram)
    are known, and not considered here.}
\end{center}
\end{figure}

In Fig.~\ref{fig::sample} we show one- and two-loop sample Feynman diagrams
contributing to process
\begin{eqnarray}
  g(p_1) g(p_2) \to Z(p_3) Z(p_4)\,,
\end{eqnarray}
where all momenta $p_i$ are incoming. The Mandelstam variables are
defined as
\begin{eqnarray}
  s &=& (p_1+p_2)^2\,,\nonumber\\
  t &=& (p_1+p_3)^2\,,\nonumber\\
  u &=& (p_1+p_4)^2\,,
\end{eqnarray}
and fulfil the property
\begin{eqnarray}
  s + t+ u &=& 2 m_Z^2\,.
\end{eqnarray}
For later convenience we also introduce the velocity $\beta$ and the transverse momentum of
the $Z$ bosons as
\begin{align}
	\beta=\sqrt{1-\frac{4m_Z^2}{s}}
	\,,\quad\quad
	p_T = \sqrt{\frac{tu-m_Z^4}{s}}
	      = \frac{\sqrt{s}}{2}\beta \sin\theta\,,
            \label{eq::pT2}
\end{align}
where $\theta$ is the scattering angle.

In this paper we consider only the top quark as the virtual particle in the loop.
We exclude from our analysis the two-loop contribution which originates from
the product of two one-loop triangle diagrams (the so-called anomaly
contribution) since this contribution is discussed in detail in
Ref.~\cite{Campbell:2016ivq}, in which exact results are presented.

The $Z$ boson has a vector and axial-vector coupling to the top quark,
for which the corresponding Feynman rule is given by
\begin{eqnarray}
  -i \frac{e}{2\sin\theta_W\cos\theta_W} \gamma^\mu 
                      \left( v_t + a_t \gamma_5 \right)
                      \,,
\end{eqnarray}
with
\begin{align}
  v_t &= \frac{1}{2}-\frac{4}{3}\sin^2\theta_W\,,\qquad
  a_t = \frac{1}{2}
          \,.
          \label{eq::vtat}
\end{align}
$\theta_W$ denotes weak mixing angle and $e=\sqrt{4\pi\alpha}$ where $\alpha$
is the fine structure constant.
The amplitude for $gg\to ZZ$ has contributions proportional to $v_t^2$ and $a_t^2$.

The polarization vectors of the gluons and $Z$ bosons are given by
$\varepsilon_{\lambda_1,\mu}(p_1)$, 
$\varepsilon_{\lambda_2,\nu}(p_2)$ and
$\varepsilon_{\lambda_3,\rho}(p_3)$,
$\varepsilon_{\lambda_4,\sigma}(p_4)$, in terms of which the amplitude can be written as
\begin{eqnarray}
\label{eq::helicity_amp}
  \mathcal{M}_{\lambda_1,\lambda_2,\lambda_3,\lambda_4} &=& A^{\mu\nu\rho\sigma}
               \varepsilon_{\lambda_1,\mu}(p_1)\varepsilon_{\lambda_2,\nu}(p_2)
               \varepsilon_{\lambda_3,\rho}(p_3)\varepsilon_{\lambda_4,\sigma}(p_4)
               \,.
\end{eqnarray}
Here the colour indices have been suppressed.
$A^{\mu\nu\rho\sigma}$ is a linear combination of 20 tensor
structures~\cite{vonManteuffel:2015msa,Caola:2015ila,phd_wellmann}
\begin{eqnarray}
  A^{\mu\nu\rho\sigma} &=& \sum_{i=1}^{20} f_i \: S_{i}^{\mu\nu\rho\sigma}
                                         \,,
  \label{eq::M1}
\end{eqnarray}
where the tensor structures $S_i$ are chosen as
\begin{alignat}{4}
  &S_{1}^{\mu\nu\rho\sigma} = g^{\mu\nu} g^{\rho\sigma}\,,\quad
  &&S_{2}^{\mu\nu\rho\sigma} = g^{\mu\rho}g^{\nu\sigma}\,,\quad
  &&S_{3}^{\mu\nu\rho\sigma} = g^{\mu\sigma} g^{\nu\rho}\,,\quad
  &&S_{4}^{\mu\nu\rho\sigma} = g^{\mu\sigma} p_1^{\rho}p_3^{\nu}\,,\quad
  \nonumber\\
  &S_{5}^{\mu\nu\rho\sigma} = g^{\mu\sigma} p_2^{\rho} p_3^{\nu}\,,\quad
  &&S_{6}^{\mu\nu\rho\sigma} = g^{\nu\sigma} p_1^{\rho} p_3^{\mu}\,,\quad
  &&S_{7}^{\mu\nu\rho\sigma} = g^{\nu\sigma} p_2^{\rho} p_3^{\mu}\,,\quad
  &&S_{8}^{\mu\nu\rho\sigma} = g^{\rho\sigma} p_3^{\mu} p_3^{\nu}\,,\quad
  \nonumber\\
  &S_{9}^{\mu\nu\rho\sigma} = g^{\mu\nu} p_1^{\rho} p_1^{\sigma}\,,\quad
  &&S_{10}^{\mu\nu\rho\sigma} = g^{\mu\nu} p_1^{\rho} p_2^{\sigma}\,,\quad
  &&S_{11}^{\mu\nu\rho\sigma} = g^{\mu\nu} p_1^{\sigma}p_2^{\rho}\,,\quad
  &&S_{12}^{\mu\nu\rho\sigma} = g^{\mu\nu} p_2^{\rho} p_2^{\sigma}\,,\quad
  \nonumber\\
  &S_{13}^{\mu\nu\rho\sigma} =  g^{\mu\rho} p_1^{\sigma} p_3^{\nu}\,,\quad
  &&S_{14}^{\mu\nu\rho\sigma} = g^{\mu\rho} p_2^{\sigma} p_3^{\nu}\,,\quad 
  &&S_{15}^{\mu\nu\rho\sigma} = g^{\nu\rho} p_1^{\sigma}  p_3^{\mu}\,,\quad
  &&S_{16}^{\mu\nu\rho\sigma} = g^{\nu\rho} p_2^{\sigma} p_3^{\mu}\,,\quad
  \nonumber\\
  &S_{17}^{\mu\nu\rho\sigma} = p_1^{\rho} p_1^{\sigma} p_3^{\mu} p_3^{\nu}\,,\quad
  &&S_{18}^{\mu\nu\rho\sigma} = p_1^{\rho} p_2^{\sigma} p_3^{\mu} p_3^{\nu}\,,\quad
  &&S_{19}^{\mu\nu\rho\sigma} = p_1^{\sigma} p_2^{\rho} p_3^{\mu} p_3^{\nu}\,,\quad
  &&S_{20}^{\mu\nu\rho\sigma} = p_2^{\rho} p_2^{\sigma} p_3^{\mu} p_3^{\nu}\,.
  \label{eq::calT}
\end{alignat}
This set omits structures which would vanish after contraction with the following choice
of polarization sums:
\begin{eqnarray}
  \sum_{\lambda_1} 
  \varepsilon_{\lambda_1,\mu}(p_1) \varepsilon_{\lambda_1,\mu^\prime}^{*}(p_1)
  &=& - g_{\mu \mu^\prime} 
  + \frac{p_{1,\mu} p_{2,\mu^\prime} + p_{2,\mu} p_{1,\mu^\prime}}{p_1\cdot p_2}
  \,,\nonumber\\
  \sum_{\lambda_2} 
  \varepsilon_{\lambda_2,\nu}(p_2) \varepsilon_{\lambda_2,\nu^\prime}^{*}(p_2)
  &=& - g_{\nu \nu^\prime} 
  + \frac{p_{2,\nu} p_{1,\nu^\prime} + p_{1,\nu} p_{2,\nu^\prime}}{p_1\cdot p_2}
  \,,\nonumber\\
  \sum_{\lambda_3} 
  \varepsilon_{\lambda_3,\rho}(p_3) \varepsilon_{\lambda_3,\rho^\prime}^{*}(p_3)
  &=& - g_{\rho \rho^\prime} + \frac{p_{3,\rho} p_{3,\rho^\prime}}{m_Z^2}
  \,,\nonumber\\
  \sum_{\lambda_4} 
  \varepsilon_{\lambda_4,\sigma}(p_4) \varepsilon_{\lambda_4,\sigma^\prime}^{*}(p_4)
  &=& - g_{\sigma \sigma^\prime} + \frac{p_{4,\sigma} p_{4,\sigma^\prime}}{m_Z^2}
  \,.
  \label{eq::pol_sum}
\end{eqnarray}
In order to obtain the scalar coefficients $ f_i$ we
construct a projector for each $S_i$ ($i=1,\ldots,20$); these are
given as linear combinations of the 138 possible rank 4 tensor structures that
can be constructed from the three independent four-vectors $p_1$, $p_2$,
$p_3$, and the metric tensor. The scalar coefficients have a perturbative expansion
in powers of the strong coupling constant which we write as
\begin{eqnarray}
  f_i
  &=& 
      \delta_{ab} \frac{\sqrt{2}G_F m_Z^2 \alpha_s(\mu)}{\pi}
      \left[ f_i^{(0)} + \frac{\alpha_s(\mu)}{\pi} f_i^{(1)} + \ldots
      \right]
      \,,
\end{eqnarray}
where $a$ and $b$ are adjoint colour indices of the incoming gluons.
The form factors $f_i^{(0)}$ and $f_i^{(1)}$ can be separated
into triangle and box contributions
\begin{eqnarray} 
 f_i^{(j)} &=& \frac{s}{3(s-m_H^2)} f_{i,\rm tri}^{(j)} 
                + f_{i,\rm box}^{(j),v_t} 
                + f_{i,\rm box}^{(j),a_t}
                \label{eq::F_tri_box}
\end{eqnarray}
where the superscripts ``$v_t$'' and ``$a_t$'' refer to the contributions
proportional to $v_t^2$ and $a_t^2$, respectively. $f_{i,\rm tri}^{(j)}$
describes contributions from diagrams which contain a Higgs-ZZ coupling,
and we note that only $f_{1,\rm tri}^{(j)}$ is non-zero.
In the case of massless quark loops,
$f_{i,\rm box}^{(j),v_t}/v_t^2 = f_{i,\rm box}^{(j),a_t}/a_t^2$~\cite{Glover:1988rg}. This property is
satisfied by the leading term of our high-energy expansions ($m_t^0 m_Z^0$) but
is violated in higher order terms, including for higher order terms in $m_Z$ since
$m_Z < m_t$.

These form factors are, at this point, divergent in 4 dimensions. We perform the
renormalization of the top quark mass, the strong coupling constant and the gluon
field to remove the ultra-violet divergences. The remaining divergences are infrared
in nature and are removed by the subtraction procedure of Ref.~\cite{Catani:1998bh},
which we outline here.

We construct finite form factors which are defined as
\begin{eqnarray}
  f_i^{(1),\rm fin} = f_i^{(1),\rm IR} - K_g^{(1)}
  f_i^{(0)} \,, 
\end{eqnarray}
where $f^{(1),\rm IR}$ is ultraviolet renormalized but still infrared
divergent.  $K_g^{(1)}$ can be found in Ref.~\cite{Catani:1998bh} and is given
by
\begin{eqnarray}
  K_g^{(1)} &=& - \left(\frac{\mu^2}{-s-i\delta}\right)^\epsilon 
  \frac{e^{\epsilon\gamma_E}}{ 2 \Gamma(1-\epsilon)}
  \left[\frac{C_A}{\epsilon^2} +
    \frac{2 \beta_0}{\epsilon}
  \right]
  \,,
  \label{eq::Ig1}
\end{eqnarray}
where $\gamma_E$ is Euler's constant.  Note that the poles in the terms
proportional to $n_f$ from Eq.~(\ref{eq::Ig1}) cancel against the counterterm
contribution induced by the $\alpha_s$ renormalization.  However, finite terms
proportional to $\log(\mu^2/(-s-i\delta))$ remain, which can be cast in the
form
\begin{eqnarray}
  f_i^{(1),\rm fin} &=& \tilde{f}_i^{(1)}
              + \beta_0 \log\left(\frac{\mu^2}{-s-i\delta}\right) f_i^{(0)}
              \,,
                                    \label{eq::Ftil}
\end{eqnarray}
with $\beta_0 = 11 C_A/12 - T n_f/3$.  Only
$\tilde{f}_i^{(1)} \equiv f_i^{(1),\rm fin}(\mu^2 = -s)$,
which are independent of $\mu$, contain new information
and thus only they will be discussed in Section~\ref{sec::NLO}.

We now discuss the work-flow for our calculation of these form factors, as expansions
in both the high-energy (Section~\ref{sec::highenergy-exp}) and large-$m_t$
(Section~\ref{sec::largemt-exp}) limits.
Analytic expressions for the results of both of these expansions can be found in the ancillary
file of this paper~\cite{progdata}. In both cases, the amplitude is generated using
\texttt{qgraf}~\cite{Nogueira:1991ex}. Each Feynman diagram is then contracted with one of the 138
possible tensor structures discussed above, as a separate computation. This splitting
is particularly important for the large-$m_t$ expansion of Section~\ref{sec::largemt-exp},
in order to avoid overly large intermediate expressions.

We additionally reproduce the exact LO result
from~\cite{Glover:1988rg} using the programs {\tt FeynArts~3.10}~\cite{Hahn:2000kx}
and {\tt FormCalc~9.8}~\cite{Hahn:2016ebn}.  The scalar Passarino-Veltman functions
$B_0$, $C_0$ and $D_0$ are rewritten in terms of polylogarithms with the help of {\tt
  Package-X}~\cite{Patel:2016fam}, which allows for a high-precision
evaluation within {\tt Mathematica}. We use this exact LO result to evaluate the
performance of our expansions and approximation methods in Section~\ref{sec::LO}.

\subsection{High-energy expansion}
\label{sec::highenergy-exp}
For each contraction we compute the fermion traces and write the result in terms of
scalar Feynman integrals, belonging to one of the integral families defined in
Refs.~\cite{Davies:2018ood,Davies:2018qvx} (there in the context of an NLO calculation
of $gg\to HH$ in the high-energy limit).
We then construct the appropriate linear combinations which are required to obtain
the form factors of the 20 tensor structures given in Eq.~(\ref{eq::calT}).
Up to this point our calculation is exact in all kinematic variables and masses.

Next, we Taylor expand both the scalar Feynman integrals and their coefficients in $m_Z$,
using the program {\tt LiteRed}~\cite{Lee:2013mka} and in-house
{\tt FORM}~\cite{Ruijl:2017dtg}
routines. For each integral family we perform an integration-by-parts (IBP) reduction
to master integrals using version~6 of {\tt FIRE}~\cite{Smirnov:2019qkx} and symmetry
relations obtained using {\tt LiteRed}~\cite{Lee:2013mka}. Since we have performed a
Taylor expansion the integrals depend on the kinematic variables and $m_t$, but no
longer on $m_Z$; this makes the IBP reduction much more tractable. For the most
complicated family (numbered~91 in Appendix~A of Ref.~\cite{Davies:2018qvx})
this takes about 4.5~days\footnote{We note that here we reduce a factor of 4
more integrals compared to Refs.~\cite{Davies:2018ood,Davies:2018qvx}. Nevertheless, the
reductions take a similar amount of CPU time due to the performance improvements of
\texttt{FIRE 6} compared to \texttt{FIRE 5.2}.}
on a 3.5~GHz machine with 32~cores.

Inserting the reduction tables into the amplitude and expanding the resulting expressions
in $m_t$ and $\epsilon$ took around three weeks on a reasonably sized cluster of computers.
Using the results for the master integrals of Refs.~\cite{Davies:2018ood,Davies:2018qvx},
we produce an expression for the amplitude expanded up to $m_t^{32}$ and $m_Z^4$.
The coefficients of the expansion terms are functions of $s$ and $t$, and are written
in terms of Harmonic Polylogarithms with a harmonic weight of at most 4, for the numerical
evaluation of which we use the package \texttt{HPL.m}~\cite{Maitre:2007kp}.

The expansions contain terms with both even and odd powers of $m_t$; the odd powers
come from the expansions of the two-loop non-planar master integrals.
Most of the odd powers cancel in the amplitude, however starting from $m_t^{3}$, odd
$m_t$ powers remain in the imaginary part of the non-abelian contribution to the
form factors. The situation is analogous to $gg\to HH$~\cite{Davies:2018qvx} where the
contributions of odd powers is discussed in detail at the level of master
integrals.

\subsection{Large-$m_t$ expansion}
\label{sec::largemt-exp}
We keep the discussion of the large-$m_t$ expansion brief, since the methods are
largely the same as used in the expansion of Higgs boson pair production, described
in detail in Ref.~\cite{Grigo:2013rya}. At the level of the individual Feynman
diagrams contracted with one of the 138 possible tensor structures, we apply an
asymptotic expansion for $m_t\gg p_1,p_2,p_3$ using the program
\texttt{exp}~\cite{Harlander:1997zb,Seidensticker:1999bb}.

This leads to one- and two-loop vacuum integrals with the scale $m_t$ multiplied
by massless three-point integrals with the scale $s$. We expand, at one
and two loops, to order $1/m_t^{12}$. After expansion, we compute the appropriate
linear combinations of the contractions in order to arrive at the coefficients of
the 20 tensor structures of Eq. (\ref{eq::M1}), yielding the large-$m_t$ expanded
expressions for the form factors defined in Eq.~(\ref{eq::F_tri_box}). For the
convenience of the reader we show the leading terms in the $1/m_t$ expansion for some
form factors in Appendix~\ref{app::example_analytic}.

In Ref.~\cite{Melnikov:2015laa} the amplitude for $gg\to ZZ$ has been
calculated at LO and NLO up to the first non-vanishing expansion term in
$1/m_t$, which only involves the axial-vector part. We find agreement after
fixing two obvious typos\footnote{In Eq.~(5) of~\cite{Melnikov:2015laa} the
  term $f^1_{\mu\rho}f^{2,\mu}_{\beta}$ should be multiplied by $(-1)$ and in
  Eq.~(7) $p_1^\mu$ and $p_2^\nu$ should be replaced by $p_2^\mu$ and
  $p_1^\nu$, respectively.}.  Furthermore, in Ref.~\cite{Campbell:2016ivq}
analytic results for the $gg\to ZZ$ amplitude projected to the triangle
contribution are presented as an expansion up to order $1/m_t^{12}$. After
performing the same projection we could successfully compare our results for
the vector and axial-vector part which constitutes a welcome check for our
approach.

\subsection{Orthogonal tensor basis}
\label{sec::orthog-tensor-basis}
The tensors given in Eq.~(\ref{eq::calT}) have the advantage of being
simple and compact. However, they are not orthogonal; this leads to non-vanishing
cross terms when squaring the amplitude.
For this reason we construct a new basis $T_i$, using the Gram-Schmidt
orthogonalization procedure, which has the property
\begin{align}
  c_i \delta_{ij} \stackrel{d\to4}{=}& \sum_{\lambda_1,\lambda_2,\lambda_3,\lambda_4}
  T_i^{\mu\nu\rho\sigma}
  T_j^{*,\mu^\prime\nu^\prime\rho^\prime\sigma^\prime} \times
  \nonumber\\
  &\varepsilon_{\lambda_1,\mu}(p_1)\varepsilon_{\lambda_2,\nu}(p_2)
  \varepsilon_{\lambda_3,\rho}(p_3)\varepsilon_{\lambda_4,\sigma}(p_4)
  \varepsilon_{\lambda_1,\mu^\prime}^{*}(p_1)\varepsilon_{\lambda_2,\nu^\prime}^{*}(p_2)
  \varepsilon_{\lambda_3,\rho^\prime}^{*}(p_3)\varepsilon_{\lambda_4,\sigma^\prime}^{*}(p_4)
  \label{eq::sum_lam}
\end{align}
For $d=4$ the coefficients $c_i$ are given by
\begin{align}
\label{eq::basis_norm}
c_{1} = c_{2} = \cdots = c_{10} = 1 
\quad , \quad
c_{11} = c_{12} = \cdots = c_{18} = p_T^2 m_Z^2
\quad , \quad
c_{19} = c_{20} = 0
\;.
\end{align}
Note that in four dimensions $c_{19}$ and $c_{20}$ vanish which means that 
in the orthogonal basis only 18 form factors contribute to the final results.
To obtain Eq.~(\ref{eq::basis_norm}) we have made use of the polarization
sums already listed in Eq.~(\ref{eq::pol_sum}).

The basis change from $S_i$ to $T_i$ is described in
Appendix~\ref{app::T}.
In terms of $T_i$ the amplitude in Eq.~(\ref{eq::M1}) reads
\begin{eqnarray}
  A^{\mu\nu\rho\sigma} &=& \sum_{i=1}^{18} F_i \frac{T^{\mu\nu\rho\sigma}_{i}}{\sqrt{c_i}}\,.
  \label{eq::M2}
\end{eqnarray}
where the factors $1/\sqrt{c_i}$ have been introduced such that the coefficients
$F_i$ are dimensionless. As for $f_i$,
the coefficients $F_i$ have a decomposition into form factors
$F_{i,\rm tri}^{(j)}$, $F_{i,\rm box}^{(j),v_t}$ and $F_{i,\rm box}^{(j),a_t}$
as described in Eq.~(\ref{eq::F_tri_box}). It is in terms of these form factors,
of the orthogonal basis of tensor structures, that we write an expression for
the differential cross section:
\begin{eqnarray}
  \frac{ {\rm d}\sigma }{ {\rm d}t } 
  &=& 
      \frac{G_F^2 m_Z^4}{ {512} \pi s^2} \left(\frac{\alpha_s}{\pi}\right)^2 
      \sum_{i=1}^{18} \left[ \left|F_i^{(0)}\right|^2
      + \frac{\alpha_s}{\pi} \left(
      {{F}_i^{(0)*}} {F}_i^{(1),\rm fin} + {F}_i^{(0)} {{F}_i^{(1),\rm fin*}} +R\right)
      \right]\,,
\end{eqnarray}
where ``$R$'' denotes the corrections  due to real radiation which we do not consider here.
The basis change is computed numerically, upon evaluation of the differential cross
section for particular values of the kinematic parameters.

\subsection{Helicity amplitudes}
\label{sec::helicity-amp}
In this subsection we describe how one can obtain the helicity amplitudes for
the process $gg\to ZZ$ from the tensor decomposition which we have introduced
above. For this purpose it is convenient to explicitly specify the
external momenta and to introduce polarization vectors as
follows:\footnote{Alternatively one can introduce the so-called
  spinor-helicity notation, see, e.g., Refs.~\cite{Dixon:2013uaa,vonManteuffel:2015msa}.}
\begin{align}
&
%%%%%%%%%%%%%
p_1=\frac{\sqrt{s}}{2}
\left(
\begin{array}{c}
1\\ 0\\ 0\\ 1
\end{array}
\right)
\!,\,\,
%%%%%%%%%%%%%
p_2=\frac{\sqrt{s}}{2}
\left(
\begin{array}{c}
1\\ 0\\ 0\\ -1
\end{array}
\right)
\!,\,\,
%%%%%%%%%%%%%
p_3=\frac{\sqrt{s}}{2}
\left(
\begin{array}{c}
-1\\ -\beta \sin \theta \\ 0\\ \beta \cos \theta 
\end{array}
\right)
\!,\,\,
%%%%%%%%%%%%%
p_4=\frac{\sqrt{s}}{2}
\left(
\begin{array}{c}
-1\\ \beta \sin \theta \\ 0\\ -\beta \cos \theta 
\end{array}
\right)
,
\nonumber\\
&
%%%%%%%%%%%%%
\varepsilon_+ (p_1)
=
\varepsilon_- (p_2)
=
[ \varepsilon_- (p_1) ]^*
=
[ \varepsilon_+ (p_2) ]^*
=
\frac{1}{\sqrt{2}}
\left(
\begin{array}{c}
0\\ i\\ 1\\ 0
\end{array}
\right)
\!,\,\,
%%%%%%%%%%%%%
\varepsilon_0 (p_3)
=
\frac{\sqrt{s}}{2m_Z}
\left(
\begin{array}{c}
\beta \\ -\sin \theta \\ 0\\ -\cos \theta
\end{array}
\right),
\nonumber\\
%%%%%%%%%%%%%
&
\varepsilon_+ (p_3)
=
\varepsilon_- (p_4)
=
[ \varepsilon_- (p_3) ]^*
=
[ \varepsilon_+ (p_4) ]^*
=
\frac{1}{\sqrt{2}}
\left(
\begin{array}{c}
0\\ i\cos \theta \\ 1\\ i \sin \theta
\end{array}
\right)
\!,\,\,
\varepsilon_0 (p_4)
=
\varepsilon_0 (p_3)\Big|_{\theta\to\theta+\pi}
\,,
%%%%%%%%%%%%%
\end{align}
where $\varepsilon_0$ denotes the
longitudinal components of polarization vectors.
Recall that all external momenta are defined as incoming and that
the polarization vectors are chosen such that
they satisfy Eq.~\eqref{eq::pol_sum}.
The helicity amplitudes $\mathcal{M}_{\lambda_1,\lambda_2,\lambda_3,\lambda_4}$
are given by Eq.~(\ref{eq::helicity_amp}).
In total there are $2\times 2\times 3\times 3=36$ helicity
amplitudes. However, due to various symmetries only eight of them are independent.
First, due to
\begin{align}
  [ p' \cdot \varepsilon _\pm (p) ]^* = p' \cdot \varepsilon _\mp (p) \,,
\end{align}
which holds for $p,p'=p_1,...,p_4$, we have
\begin{align}
  | \mathcal{M}_{-\lambda_1,-\lambda_2,-\lambda_3,-\lambda_4} |  =
  | \mathcal{M}_{\lambda_1,\lambda_2,\lambda_3,\lambda_4} |  \,,
  \label{eq:sym1}
\end{align}
which reduces the number of independent amplitudes to 18.  Furthermore, there
are additional symmetries~\cite{Glover:1988rg}
relating helicity amplitudes with
different polarization states
\begin{alignat}{2}
  \mathcal{M}_{+++-} &= &&\mathcal{M}_{++-+}\,,\nonumber\\
  \mathcal{M}_{+---} &= &&\mathcal{M}_{+-++}\,,\nonumber\\
  \mathcal{M}_{++\pm 0} &= &&\mathcal{M}_{++0\pm }\,,\nonumber\\
  \mathcal{M}_{+-\pm 0} &= - &&\mathcal{M}_{+-0\mp }\,,
                          \label{eq:sym2}
\end{alignat}
and there are symmetry relations due to $\beta \rightarrow -\beta$,
\begin{align}
  \mathcal{M}_{++--} & = \mathcal{M}_{++++}\Big|_{\beta \rightarrow -\beta}\,,\nonumber\\
  \mathcal{M}_{+--+} & = \mathcal{M}_{+-+-}\Big|_{\beta \rightarrow -\beta}\,,\nonumber\\
  \mathcal{M}_{+\pm +0} & = \mathcal{M}_{+\pm -0}\Big|_{\beta \rightarrow -\beta}\,.
                          \label{eq:sym3}
\end{align}
Note that this replacement changes none of the Mandelstam variables $s,t,u$.
Using Eqs.~\eqref{eq:sym2} and~\eqref{eq:sym3} reduces the number of
independent helicity amplitudes by six and four, respectively,
and we arrive at eight independent helicity amplitudes.

It turns out that the above symmetries are fulfilled
when the form factors satisfies the relations
\begin{align}
f_{12}=f_9\,,\quad
f_{20}=f_{17}\,,\quad
f_{16}=-f_4\,,\quad
f_{15}=-f_5\,,\quad
f_{14}=-f_6\,,\quad
f_{13}=-f_7\,.
\label{eq:sym-ff}
\end{align}
Note that up to this point
we do not make use of any approximation.
We use the relations~\eqref{eq:sym-ff}
as a cross check of our calculations.

The LO results for the eight independent helicity amplitudes
are provided in Ref.\cite{Glover:1988rg} and we confirm the agreement between them
and our results.\footnote{
Note that in Ref.\cite{Glover:1988rg}
one has to replace the $D$-functions (but not the $B$- or $C$-functions)
with $D\to i\pi^2 D$ in order to obtain the correct results.
Additionally, a factor $1/3$ is missing for the contributions 
from the triangle diagrams in the large-$m_t$ limit.
}
For the results of the high-energy expansion we have expanded in
$m_Z$, making the symmetry relations due to $\beta \rightarrow -\beta$ hard
to realize, since $\beta = 1 -2m_Z^2/s + \mathcal{O}(m_Z^4)$ and we do not
distinguish the origin of $m_Z$ terms in the expression. For this reason,
in the ancillary file \cite{progdata} we provide results for the twelve helicity
amplitudes
\begin{alignat}{3}
&\mathcal{M}_{++++}\:,\quad &&\mathcal{M}_{++--}\:,\quad &&\mathcal{M}_{+-+-}\:, \nonumber\\
&\mathcal{M}_{+--+}\:,\quad &&\mathcal{M}_{+++0}\:,\quad &&\mathcal{M}_{++-0}\:, \nonumber\\
&\mathcal{M}_{+-+0}\:,\quad &&\mathcal{M}_{+--0}\:,\quad &&\mathcal{M}_{+++-}\:, \nonumber\\
&\mathcal{M}_{++00}\:,\quad &&\mathcal{M}_{+-++}\:,\quad &&\mathcal{M}_{+-00}\:.
\end{alignat}

%- }}}

%- {{{ LO results and Pade improvement:

\section{\label{sec::LO}Comparison at leading order}

This section is devoted to the discussion of the LO contribution to $gg\to ZZ$
with virtual top quarks. We first consider the form factors and helicity amplitudes,
and compare the
large-$m_t$ and high-energy expansions with the exact results. Afterwards we discuss our
approach to improve the radius of convergence of our expansions, which is
based on Pad\'e approximations.  We furthermore investigate the importance of
finite $Z$ boson mass corrections.
For the numerical evaluation we use the following input
values~\cite{Tanabashi:2018oca} 
\begin{eqnarray}
  G_F &=& 1.1663787 \times 10^{-5}~\mbox{GeV}^{-2} \,,\nonumber\\
  \sin^2\theta_W &=& 0.23122\,,\nonumber\\
  \alpha_s(m_Z) &=& 0.1181\,,\nonumber\\
  m_Z &=& 91.1876~\mbox{GeV} \,,\nonumber\\
  m_H &=& 125.10~\mbox{GeV} \,,\nonumber\\
  m_t &=& 172.9~\mbox{GeV} \,.
\end{eqnarray}

\begin{figure}[b]
  \includegraphics[width=\textwidth]{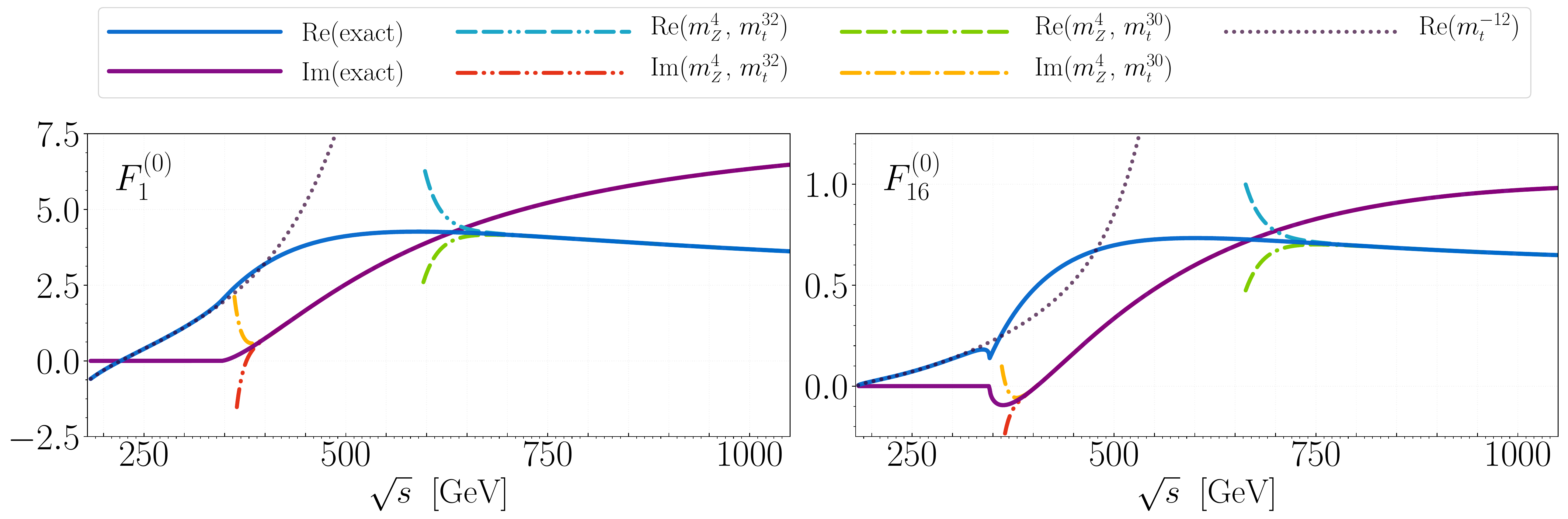}
  \caption{\label{fig::T_LO}The LO form factors of the tensor structure $T_1$ and
    $T_{16}$ as a function of $\sqrt{s}$, plotted for $\theta = \pi/2$.
    Both the real and imaginary parts are shown. Solid, dash-dotted and
    dotted lines correspond to the exact, high-energy and large-$m_t$ results.}
\end{figure}

As typical examples for the LO form factors, in Fig.~\ref{fig::T_LO} we show
the results for $F_1^{(0)}$ and $F_{16}^{(0)}$ as a function of the partonic
center-of-mass energy $\sqrt{s}$.  For the scattering angle we choose
$\theta = \pi/2$.  The solid blue and purple lines correspond to the real and
imaginary parts of the exact result. The dotted curve includes seven terms (up
to $1/m_t^{12}$) in the large-$m_t$ expansion and agrees with the blue curve
almost up to the top quark threshold at $\sqrt{s}\approx 2m_t$.
The dash-dotted lines correspond to the high-energy expansion. Both for the real
and imaginary parts we plot the expansions including terms up to $m_t^{30}$ and
$m_t^{32}$. One observes that they start to deviate from the exact result around
the same value of $\sqrt{s}$.
In fact, in these plots it is sufficient to include expansion terms only up to
$m_t^{16}$ to have a very similar high-energy approximation.
Thus, the high-energy expansions
approximate the exact curves well for $\sqrt{s}$ values above about 750~GeV and
400~GeV for the real and imaginary parts, respectively. Similar plots for all
20 form factors are shown in Appendix~\ref{app::T_LO}.

\begin{figure}[t]
  \includegraphics[width=\textwidth]{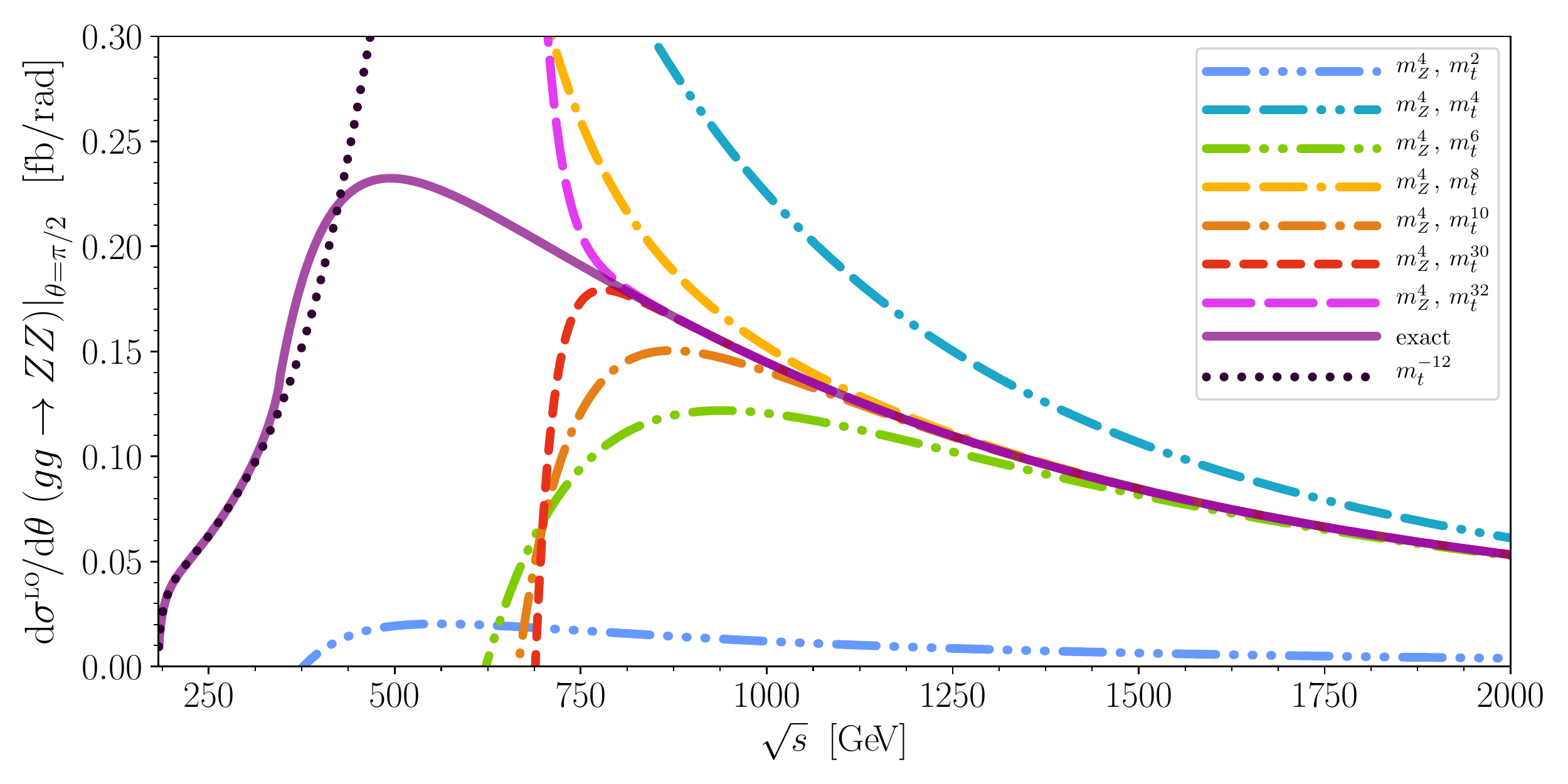}
  \caption{\label{fig::dsig_LO}LO partonic differential cross section for
    $gg\to ZZ$ for $\theta = \pi/2$.  Solid, dash-dotted and dotted lines
    correspond to the exact, high-energy and large-$m_t$ results.}
\end{figure}

Fig.~\ref{fig::dsig_LO} shows (again for $\theta = \pi/2$) the LO partonic
cross section as a function of $\sqrt{s}$.  For low values of $\sqrt{s}$ we
observe that the large-$m_t$ result (dotted) approximates the exact curve
(solid) well, almost up to the top quark threshold. The remaining curves
(dashed and dash-dotted) incorporate results from the high-energy
expansion.
We show a selection of expansion depths between $m_t^{2}$ and $m_t^{32}$.
One observes that five to six expansion terms are necessary in order to obtain a
good approximation of the exact result for $\sqrt{s}\gtrsim 1000$~GeV. The
deeper expansion depths show agreement down to $\sqrt{s} \approx 750$~GeV,
which cannot be further improved even by including terms up to $m_t^{32}$.
It appears that the simple expansions in $m_t^2/s$, $m_t^2/t$ and
$m_t^2/u$ have a finite radius of convergence, which for $\theta = \pi/2$
manifests itself around $\sqrt{s} \approx 750$~GeV.  This feature can be
understood by inspecting the functions which are present in the exact one-loop
result. Among others we have identified logarithms and di-logarithms which
depend on the quantity
\begin{align}
  X\:=\:\sqrt{ 1 + \frac{4 s m_t^2} {ut-m_Z^4} }
  \:=\:
  \sqrt{1+\frac{4sm_t^2}{ut \vphantom{m_Z^4}}} + \mathcal{O}\left(m_Z^4\right)
  \,
\end{align}
which has, in the high-energy limit, a radius of convergence of $ut/s = 4m_t^2$.
For $\theta=\pi/2$ we have $t=u=-s/2$
which leads to $\sqrt{s}=4m_t\simeq 700~\mathrm{GeV}$.

\begin{figure}[t]
  \includegraphics[width=\textwidth]{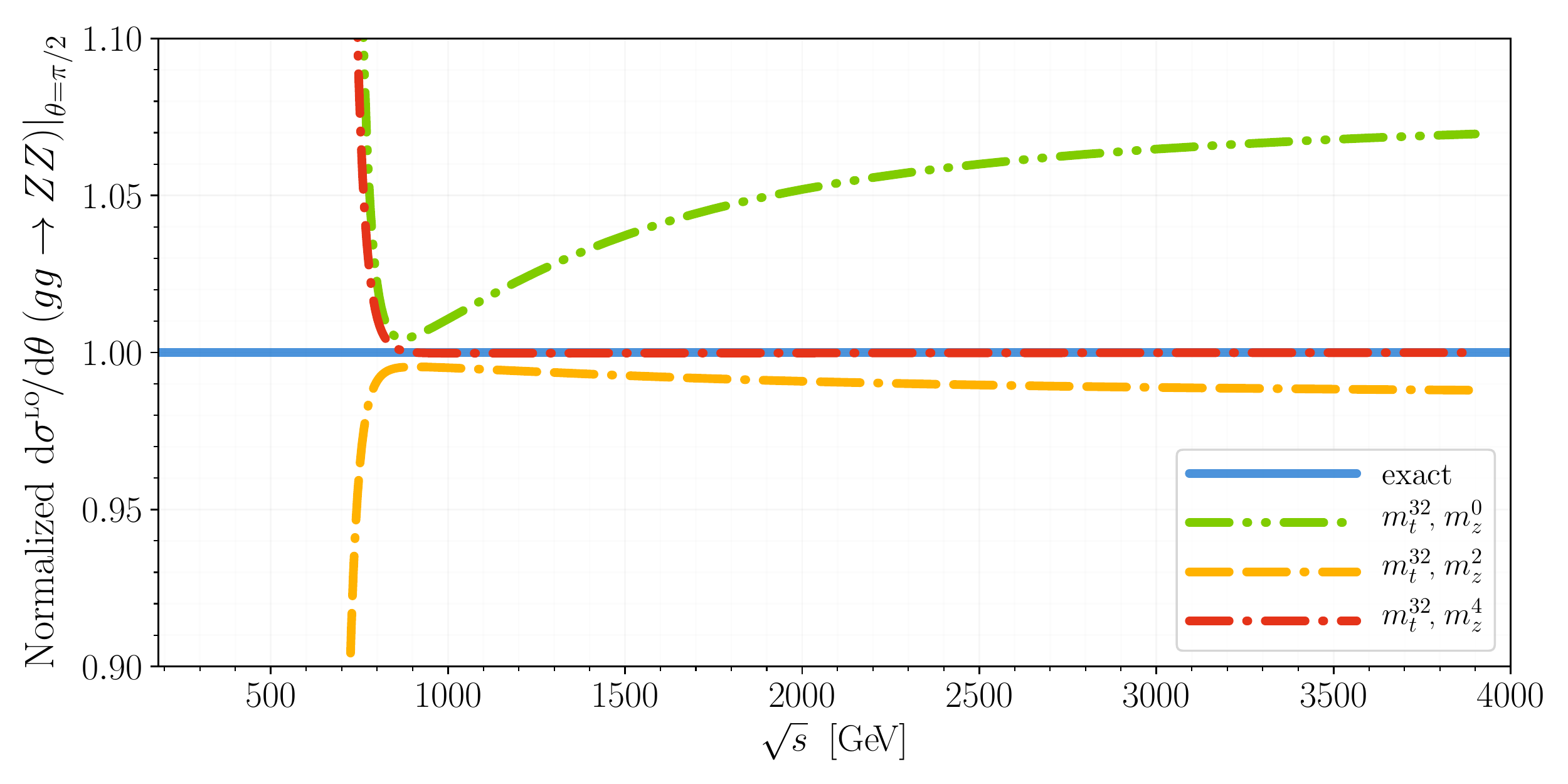}
  \caption{\label{fig::dsig_LO_mz}LO partonic differential cross section for
    $gg\to ZZ$ for $\theta = \pi/2$, normalized to the exact result.
    The three high-energy expansions contain terms up to order 
    $m_t^{32}$ and $m_Z^0$, $m_Z^2$ and $m_Z^4$.}
\end{figure}

Let us next discuss the importance of finite $m_Z$ terms.  The high-energy
approximations shown in Figs.~\ref{fig::T_LO} and~\ref{fig::dsig_LO} include
terms up to order $m_Z^4$, i.e., three expansion terms.
In Fig.~\ref{fig::dsig_LO_mz} we show how the number of expansion terms in
$m_Z^4$ affects the quality of the expansion of the LO differential cross
section. Curves including terms to $m_t^{32}$ and $m_Z^0$, $m_Z^2$ and
$m_Z^4$ are shown, normalized to the exact result. For all three curves we
observe, as discussed above, a divergent behaviour for $\sqrt{s} \lesssim 750$~GeV.
The $m_Z^0$ curve shows a more than 5\% deviation from the exact result and
including the $m_Z^2$ term leads to a significant improvement, with the deviation
reducing to around 1\%. Finally, including the $m_Z^4$ term produces a per-mille
level agreement with the exact result, which motivates our computation of the
$m_Z^4$ expansion terms of the NLO quantities discussed in later sections of this
paper.

%- }}}

%- {{{ Pade:

\section{\label{sec::pade}Pad\'e-improvement of the high energy expansion}
In Section~\ref{sec::LO} we investigated the behaviour of the expansions and,
in particular, noted that the high-energy expansion fails to converge below
$\sqrt{s}\approx 750$GeV regardless of how many expansion terms are included.
In this section we discuss a method by which we can extend the prediction of the
high-energy expansion to smaller values of $\sqrt{s}$.

The method is an extended version of the approach used in Ref.~\cite{Davies:2019dfy}
in the context of Higgs boson pair production, and we describe it in detail below.
It is based on the construction of a number of Pad\'e approximants using the terms
of the high-energy expansion, and subsequently combining the approximants to
produce a central value and uncertainty estimate for a given phase-space point
$\{\sqrt{s}, p_T\}$. We describe the procedure in terms of a generic quantity
${\cal F}$ for which we assume an expansion in $m_t$ is available.
${\cal F}$ also depends on the kinematic quantities $s$ and $p_T$ and on $m_Z$.
In our practical applications ${\cal F}$ can be either a form factor, a helicity
amplitude or the virtual finite cross section defined in Section~\ref{sec::NLO}.

The approximation procedure for ${\cal F}$ is then as follows:
\begin{itemize}
\item We write ${\cal F}$ as an expansion in $m_t$ and define
  \begin{eqnarray}
    {\cal F}^N &=& {\cal F}_{0}+\sum_{i=2}^{N} {\cal F}_{i}\, m_t^i
                             \,,
                             \label{eq::Vfin_exp}
  \end{eqnarray}
  where ${\cal F}_{0}$ contains the exact (in $m_t$ and $m_Z$) expressions
  of the LO contributions. ${\cal F}_{i}$ are the $m_t$ expansion coefficients.

\item
  We apply the replacements $m_t^{2k} \to m_t^{2k} x^k$
  and $m_t^{2k-1} \to m_t^{2k-1} x^k$ for the odd and even powers of $m_t$.
  We insert numerical values for $m_t$, $m_Z$, $s$ and $p_T$, yielding a
  polynomial in the variable $x$.

\item
  Next we construct Pad\'e approximants of ${\cal F}^N$ in
  the variable $x$ and write ${\cal F}^N$ 
  as a rational function of the form
  \begin{eqnarray}
    {\cal F}^N
    &=& 
        \frac{a_0 + a_1 x + \ldots + a_n x^n}{1 + b_1 x + \ldots + b_m x^m}
          \,\, \equiv\,\, [n/m](x)\,.
          \label{eq::Pade}
  \end{eqnarray}
  The coefficients $a_i$ and $b_i$ are determined by comparing the coefficients of $x^k$
  after expanding the right-hand side of Eq.~(\ref{eq::Pade}) around the point $x=0$.
  Evaluation of this rational function at $x=1$ yields the Pad\'e approximated value
  of ${\cal F}^N$.
\end{itemize}

The numerator and denominator degrees in Eq.~(\ref{eq::Pade}) are free
parameters; one only must ensure that $n+m\le N/2$ such that a sufficient
number of expansions terms are available to determine the coefficients $a_i$ and
$b_i$. We construct many Pad\'e approximations
and combine them to obtain a prediction for the central
value and the uncertainty of ${\cal F}$.

The rational function of Eq.~(\ref{eq::Pade}) develops poles
which, for some Pad\'e approximants, might lie close to the evaluation point $x=1$
and yield unphysical results.
In the following we describe a weighting scheme which minimizes
the influence of such Pad\'e approximants. We call this approach a {\it pole distance
reweighted} (PDR) Pad\'e approximation.
\begin{itemize}
  \item For each phase-space point $\{\sqrt{s},p_T\}$ we compute, for each Pad\'e
    approximant, the value at $x=1$ and the distance of the nearest pole which
    we denote by $\alpha_i$ and $\beta_i$, respectively.

  \item Introduce a weighting function, which reduces the impact of values $\alpha_i$
  from Pad\'e approximations with poles close to $x=1$. We define
    \begin{eqnarray}
      \omega_{i,\rm poles} &=& \frac{\beta_i^2}{\sum_j\beta_j^2}\,,
                   \label{eq::w_poles}
    \end{eqnarray}
    where the sum runs over all Pad\'e approximants under consideration.

  \item Use the values $\alpha_i$ and $\omega_{i,\rm poles}$ to compute the weighted
  average and weighted standard deviation of the Pad\'e approximants,
    \begin{eqnarray}
      \alpha &=& \sum_i\omega_{i,\rm poles}\alpha_i \,,
                 \qquad
      \delta_{\alpha} = \sqrt{\frac{\sum_i\omega_{i,\rm
                          poles}\left(\alpha_i-\alpha\right)^2}{1-\sum_i\omega_{i,\rm
                          poles}^2}}
      \,.
                 \label{eq::alpha_delta}
    \end{eqnarray}
    These form the central value and error estimate of the approximation.

  \end{itemize}

At this point, the procedure is the same as that of Ref.~\cite{Davies:2019dfy}, in
which expansions up to $m_t^{30}$ and $m_t^{32}$ were used to create Pad\'e approximants
with $15\le n+m\le 16$, with the additional restriction to ``near-diagonal'' approximants
which satisfy $|n-m|\le2$. This results in 5 possible approximants,
\begin{eqnarray}
	\left\lbrace[7/8],[8/7],[7/9],[8/8],[9/7] \right\rbrace\,,
	\label{eq::old_pade_set}
\end{eqnarray}
which were weighted according
to the above procedure to produce a central value and error estimate.

In this paper we further refine the method which allows us to loosen the restrictions
and thus include more approximants in the computation. We introduce two additional
weights into the averaging procedure which a) emphasize the contribution from Pad\'e
approximants which are derived from a larger number of expansion terms and b) emphasize
the contribution from ``near-diagonal'' approximants. These weights are defined as follows:
\begin{enumerate}
\renewcommand{\theenumi}{\alph{enumi})}
	\item An $[n_i/m_i]$ Pad\'e approximant is weighted by
		\begin{eqnarray}
			\omega_{i,\rm input} &=& \frac{(n_i+m_i)^2}{\sum_j (n_j+m_j)^2}\,.
            \label{eq::w_input}
		\end{eqnarray}
	\item An $[n_i/m_i]$ Pad\'e approximant is also weighted by
		\begin{eqnarray}
			\omega_{i,\rm diag} &=& \frac{|n_i-m_i|^2}{\sum_j |n_j-m_j|^2}\,.
			\label{eq::w_diag}
		\end{eqnarray}
\end{enumerate}
As above, the sums run over all Pad\'e approximants under consideration.
The weights of Eqs. (\ref{eq::w_poles}), (\ref{eq::w_input}) and (\ref{eq::w_diag}) are
combined according to
\begin{eqnarray}
	\omega_i &=& 
	\frac{ \left[ \omega_{i,\rm poles}\cdot \omega_{i,\rm input}\cdot (1-\omega_{i,\rm diag}) \right]^2 }
	{\sum_j \left[ \omega_{j,\rm poles} \cdot
	\omega_{j,\rm input} \cdot (1-\omega_{j,\rm diag}) \right]^2 }\,,
\end{eqnarray}
and used to form a central value and error estimate
\begin{eqnarray}
	\alpha &=& \sum_i\omega_{i}\alpha_i \,, \qquad
	\delta_{\alpha} = \sqrt{\frac{\sum_i\omega_{i}\left(\alpha_i-\alpha\right)^2}
		{1-\sum_i\omega_{i}^2}}\,.
	\label{eq::alpha_delta_new}
\end{eqnarray}
The approximation of Eq.~(\ref{eq::alpha_delta}) used in Ref.~\cite{Davies:2019dfy},
with the restrictions described above, can be considered to be a special case of the
same procedure with the weights of Eqs.~(\ref{eq::w_input}) and (\ref{eq::w_diag})
replaced with step functions. In this refined procedure we include a wider set of
Pad\'e approximants. We define the quantities $N_{\rm low}$ and $N_{\rm high}$
such that
\begin{align}
	N_{\rm low}\le n+m \le N_{\rm high}\quad
	\textnormal{and}\quad N_{\rm low} \le n + m - | n - m | \,.
	\label{eq::N_low_high}
\end{align}
In Section~\ref{sec::NLO} we will study the quality of the approximations due to
the choices of $\{N_{\rm low}, N_{\rm high}\} = \{10,16\}$, $\{9,13\}$, $\{7,11\}$
and $\{5,9\}$. The best approximation is given by $\{10,16\}$ which contains the
following Pad\'e approximants,
\begin{align*}
\Big\{
& [5/5], [5/6], [6/5], [5/7], [7/5], [6/6], [5/8],
 [8/5], [6/7], [7/6], [5/9], [9/5], [6/8], [8/6], [7/7],\\
& [5/10], [10/5], [6/9], [9/6], [5/11], [11/5],
 [7/8], [8/7], [6/10], [10/6], [7/9], [9/7], [8/8]
\Big\}\,,
\end{align*}
a much larger number compared to the method of Ref.~\cite{Davies:2019dfy},
listed in Eq.~(\ref{eq::old_pade_set}).

\begin{figure}[t]
  \begin{tabular}{cc}
	\hspace{-4mm}
    \includegraphics[width=0.49\textwidth]{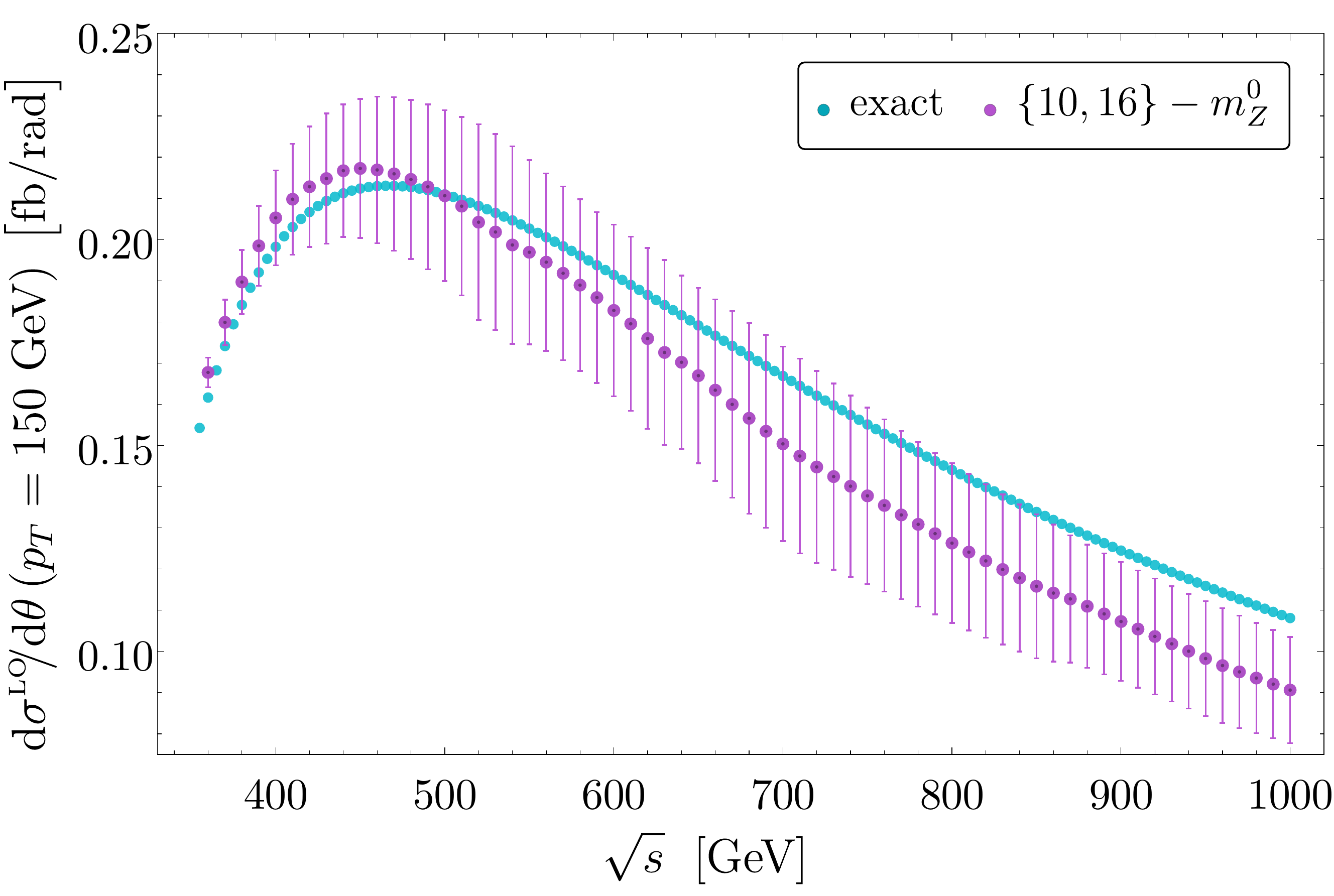}
    &
    \includegraphics[width=0.49\textwidth]{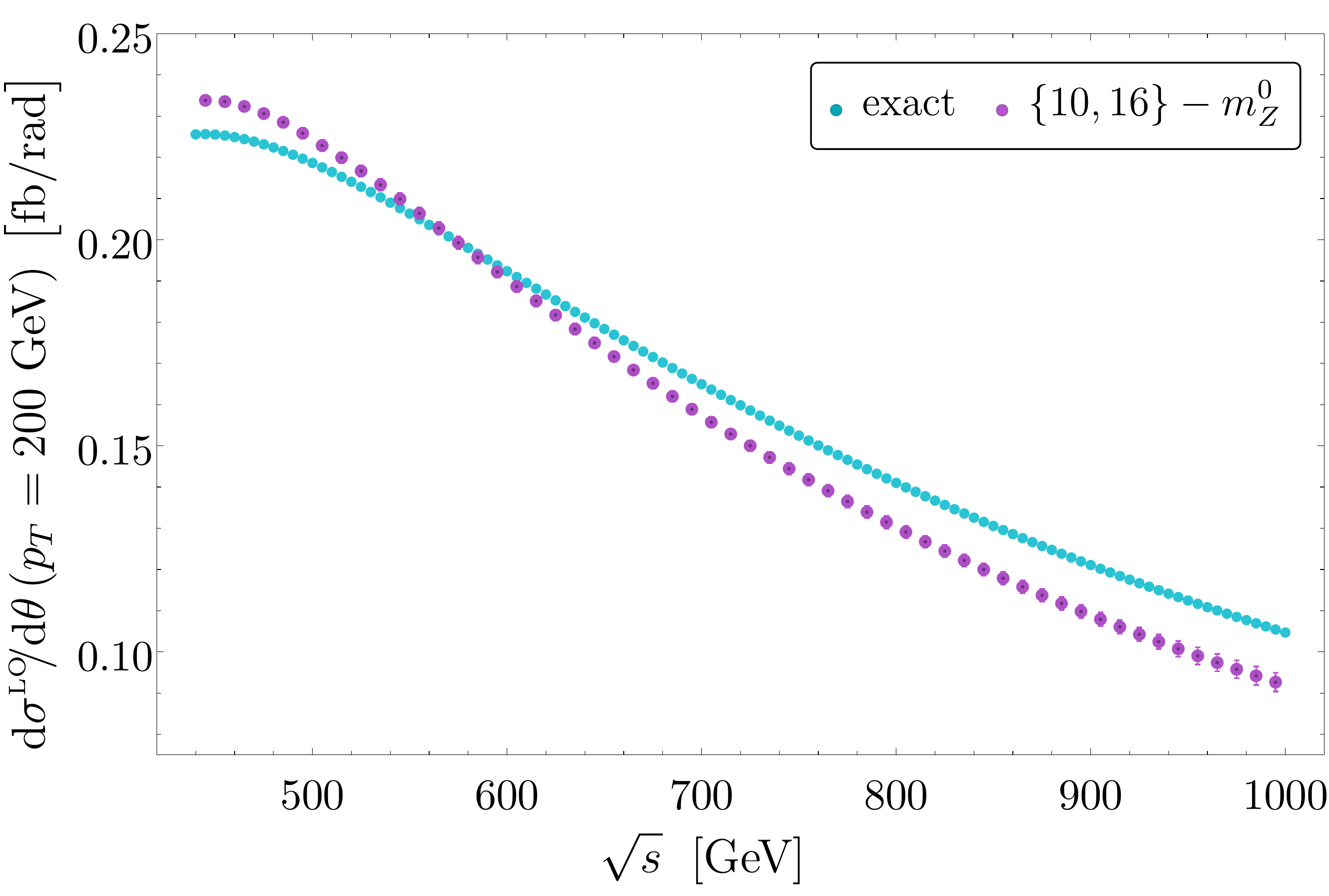}
    \\
	\hspace{-4mm}
    \includegraphics[width=0.49\textwidth]{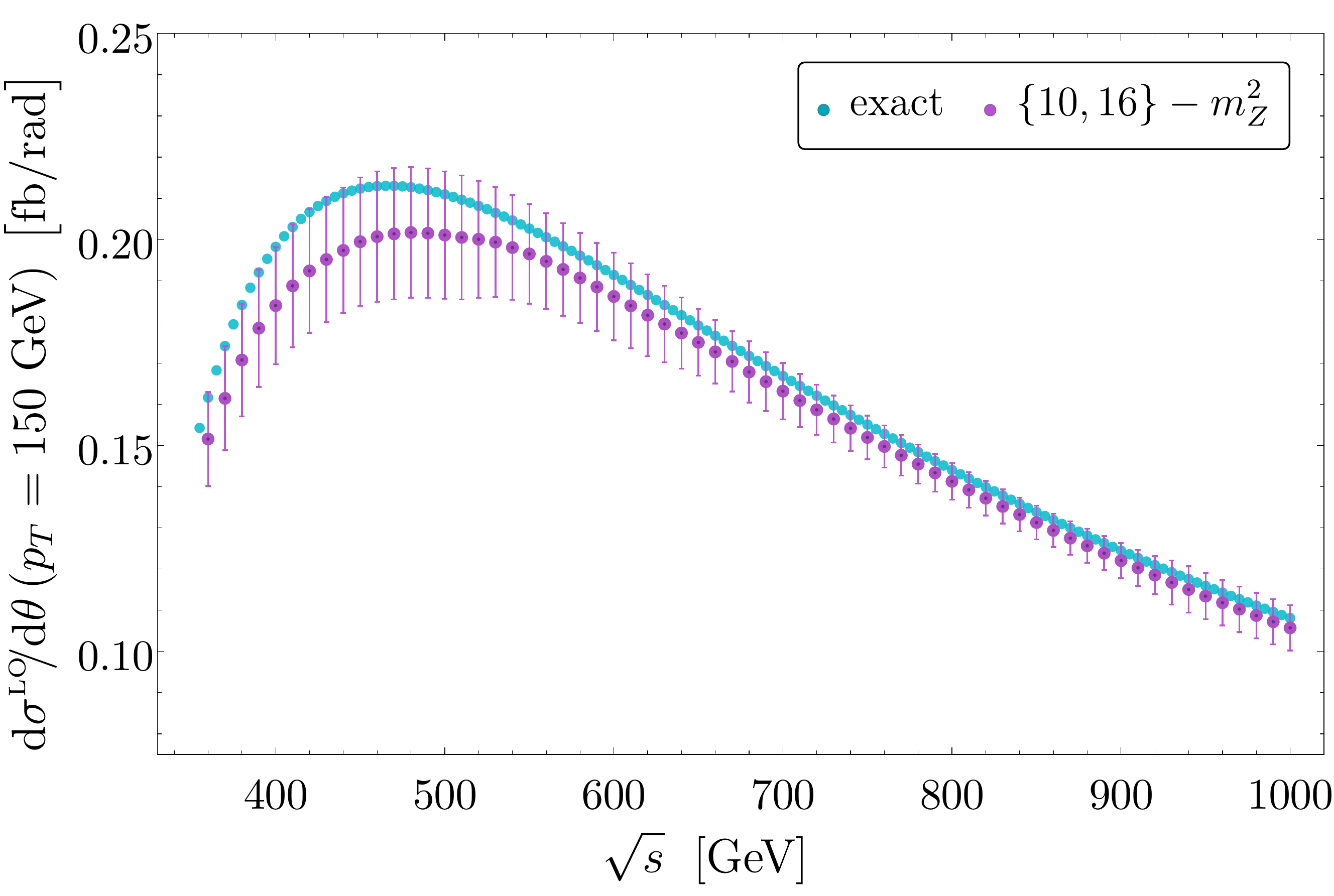}
    &
    \includegraphics[width=0.49\textwidth]{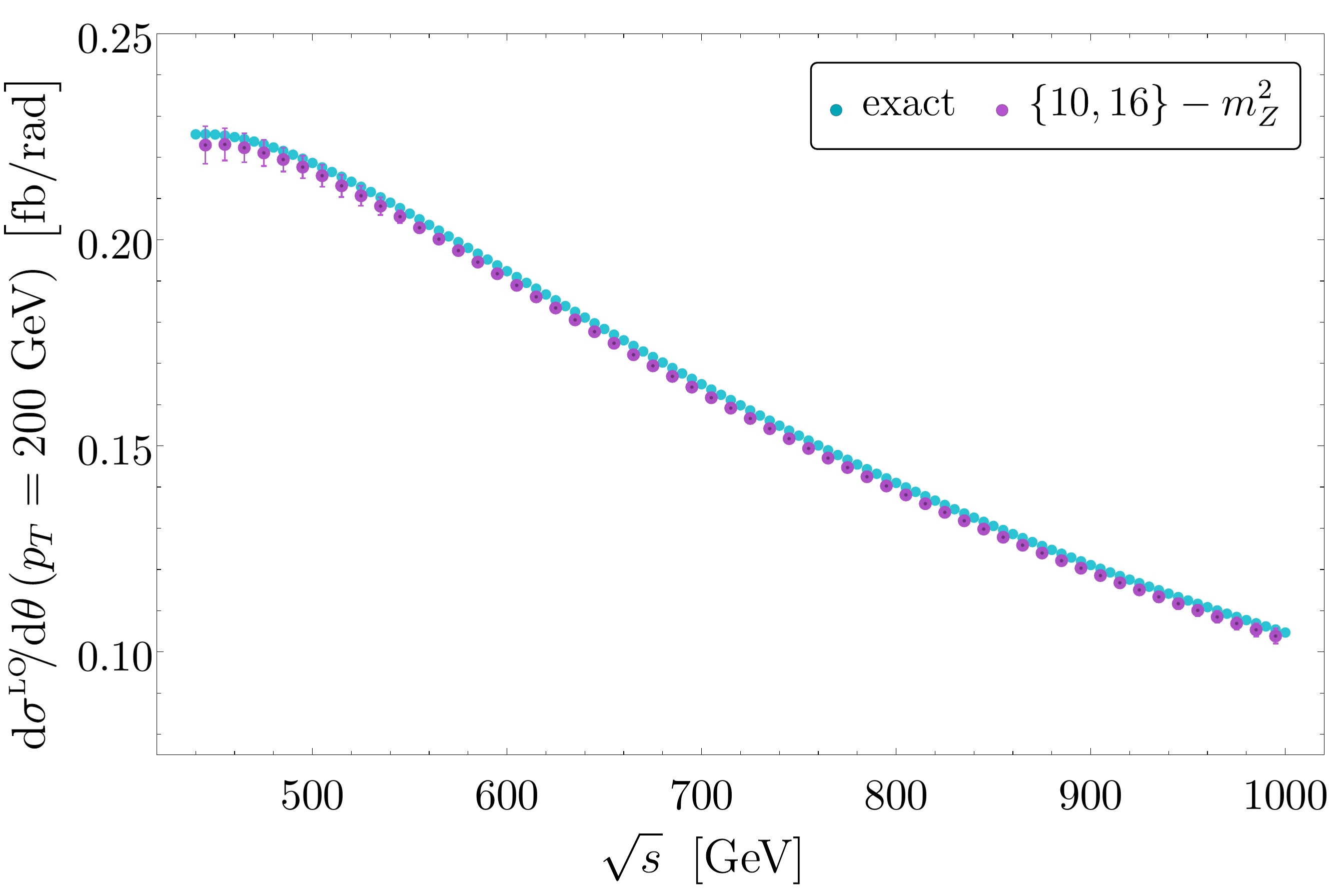}
    \\
	\hspace{-4mm}
    \includegraphics[width=0.49\textwidth]{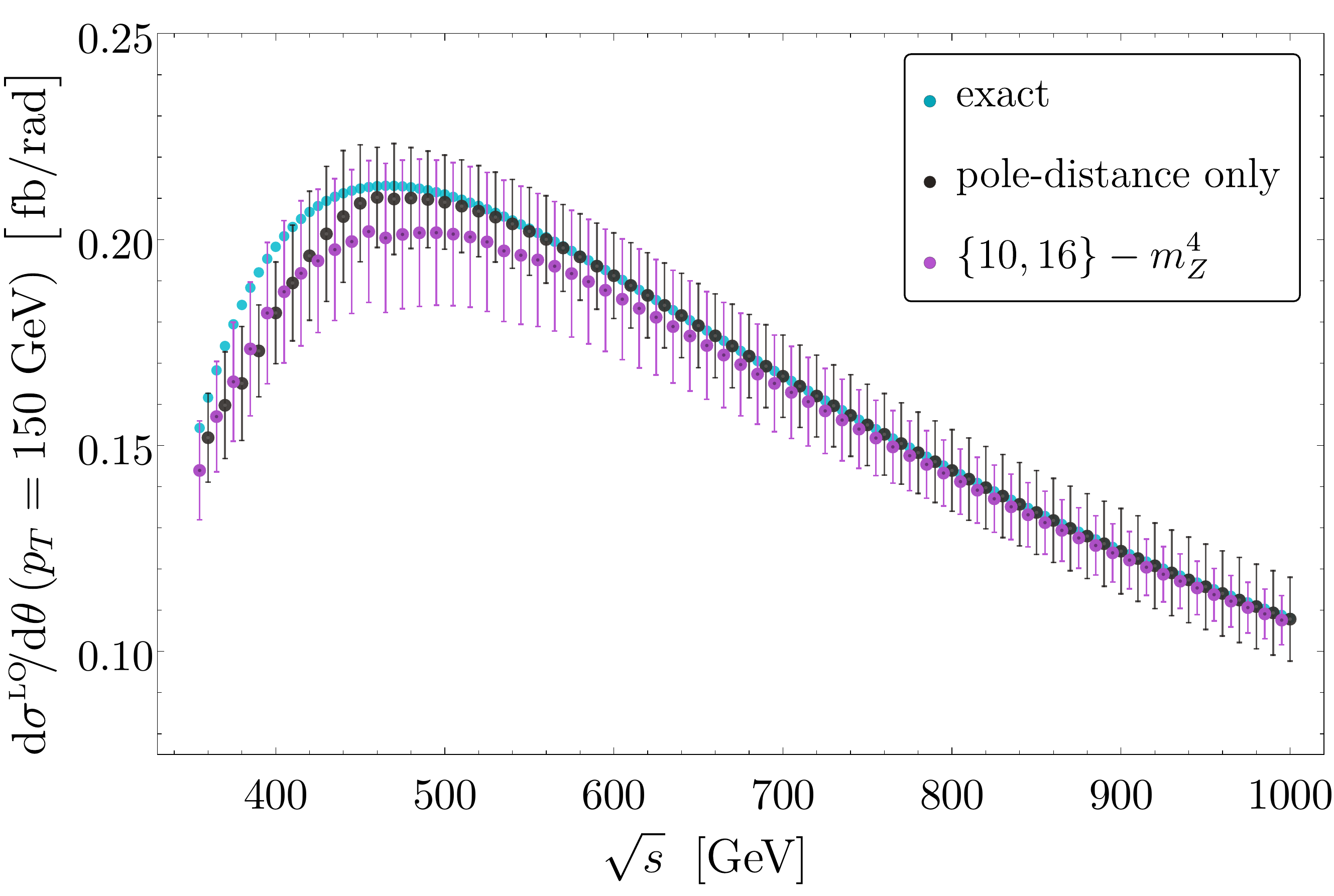}
    &
    \includegraphics[width=0.49\textwidth]{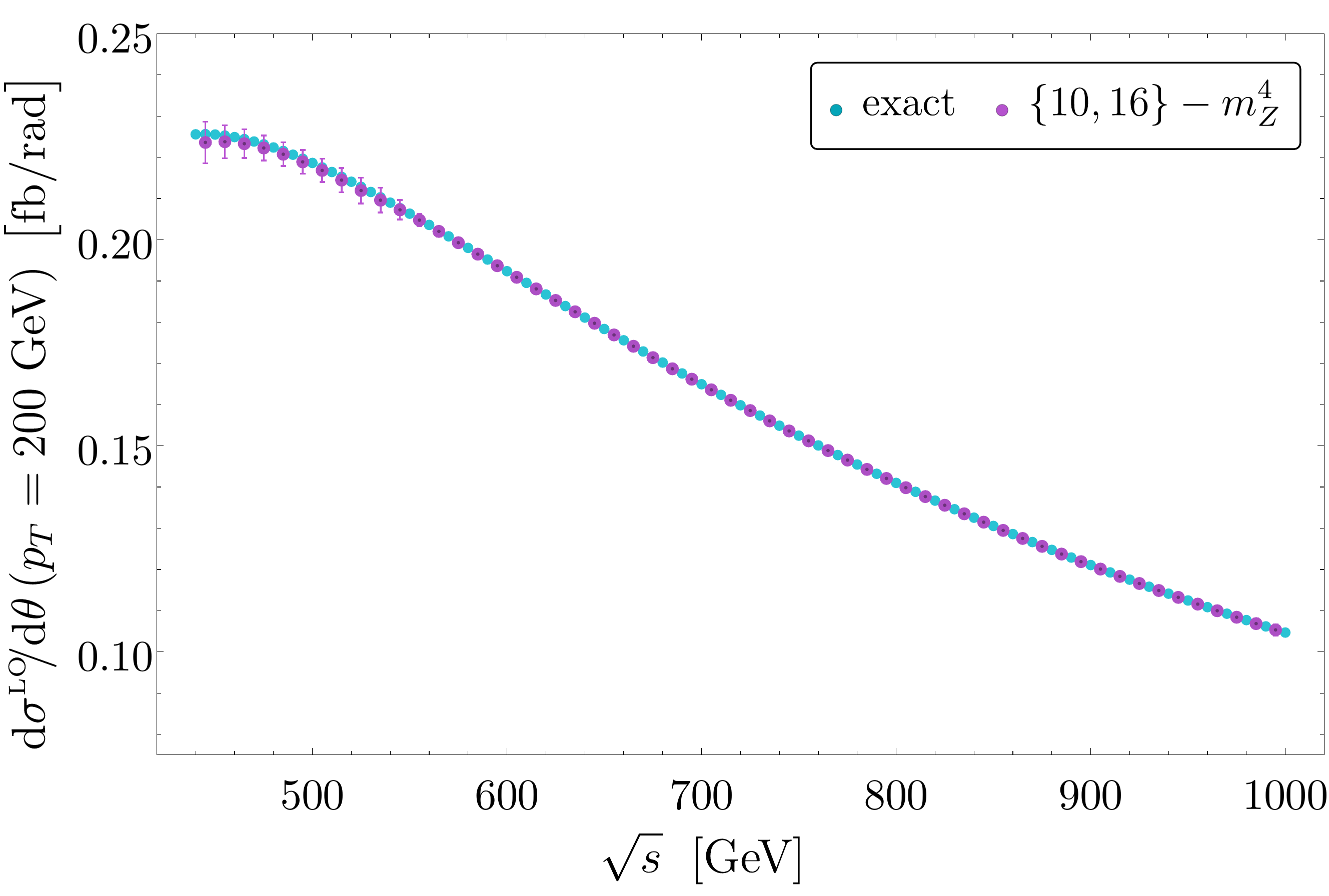}
  \end{tabular}
  \caption{\label{fig::dsig_LO_pade}LO partonic differential cross section for
    $gg\to ZZ$ for $p_T=150$~GeV (left column) and $p_T=200$~GeV (right column) as a
    function of $\sqrt{s}$.
    The blue points are the exact result, and the purple
    points are the central values with uncertainties according to the prescription
    of Eq.~(\ref{eq::alpha_delta_new}). The bottom left plot shows additionally
    central values and uncertainties according to Eq.~(\ref{eq::alpha_delta}).
    The first, second and third rows show Pad\'e approximants constructed from
    expansions including terms to $m_Z^0$, $m_Z^2$ and $m_Z^4$, respectively.}
\end{figure}

In Fig.~\ref{fig::dsig_LO_pade} we demonstrate the effect of including higher
order terms in the expansion in $m_Z$ in the construction of the Pad\'e
approximants for the LO differential cross section ${\rm d}\sigma/{\rm d}\theta$.
We show plots for $p_T = 150$~GeV and $p_T = 200$~GeV and, from
top to bottom, approximations formed from high-energy expansions which include
$m_Z^0$, $m_Z^2$ and $m_Z^4$ terms.
One observes that it is crucial to include corrections at least to order $m_Z^2$,
and that the results are further improved by including additionally the $m_Z^4$ terms.
These improvements are in line with the expectations due to the behaviour of the
high-energy expansion as demonstrated in Fig.~\ref{fig::dsig_LO_mz}.
We note that in the $p_T = 200$~GeV plots, the majority of the points have error
bars which are too small to be visible. After including the higher order $m_Z$ terms,
the exact results lie within the error estimates of the approximations, demonstrating
that they are realistic.

The bottom-left plot of Fig.~\ref{fig::dsig_LO_pade} we additionally show, in black,
a Pad\'e approximation according to the simpler prescription of
Eq.~(\ref{eq::alpha_delta}) using the five Pad\'e approximants of the set
Eq.~(\ref{eq::old_pade_set}).
One observes that for small values of $\sqrt{s}$ the
exact result lies outside of the error estimates and that for large values of
$\sqrt{s}$ the errors appear to be overestimated. In our view the purple points
provide a more reasonable description of the uncertainty.

It is interesting to have a closer look at the effective expansion parameters
entering the high-energy expansion; the final result is expressed as an
expansion in $m_t^2/s$, $m_t^2/t$ and $m_t^2/u$. Fig.~\ref{fig::dsig_LO_pade}
shows that the Pad\'{e}-improved approximations reproduce the exact result
for rather low values of $p_T$ and $\sqrt{s}$, such as
$\{p_T,\sqrt{s}\} = \{200,450\}$~GeV or $\{p_T,\sqrt{s}\} = \{150,500\}$.
For these points, the expansion parameters $\{m_t^2/s,m_t^2/t,m_t^2/u\}$
have the values $\{0.15,-0.42,-0.26\}$ and $\{0.12,-1.08,-0.14\}$. In both
cases one parameter becomes large or even exceeds $1$, however the Pad\'e
approximants nonetheless produce reliable results.

%- }}}
%- {{{ NLO results for $gg\to ZZ$:

\clearpage

\section{\label{sec::NLO}NLO results for $gg\to ZZ$ with virtual top quarks}
\begin{figure}[b]
  \includegraphics[width=\textwidth]{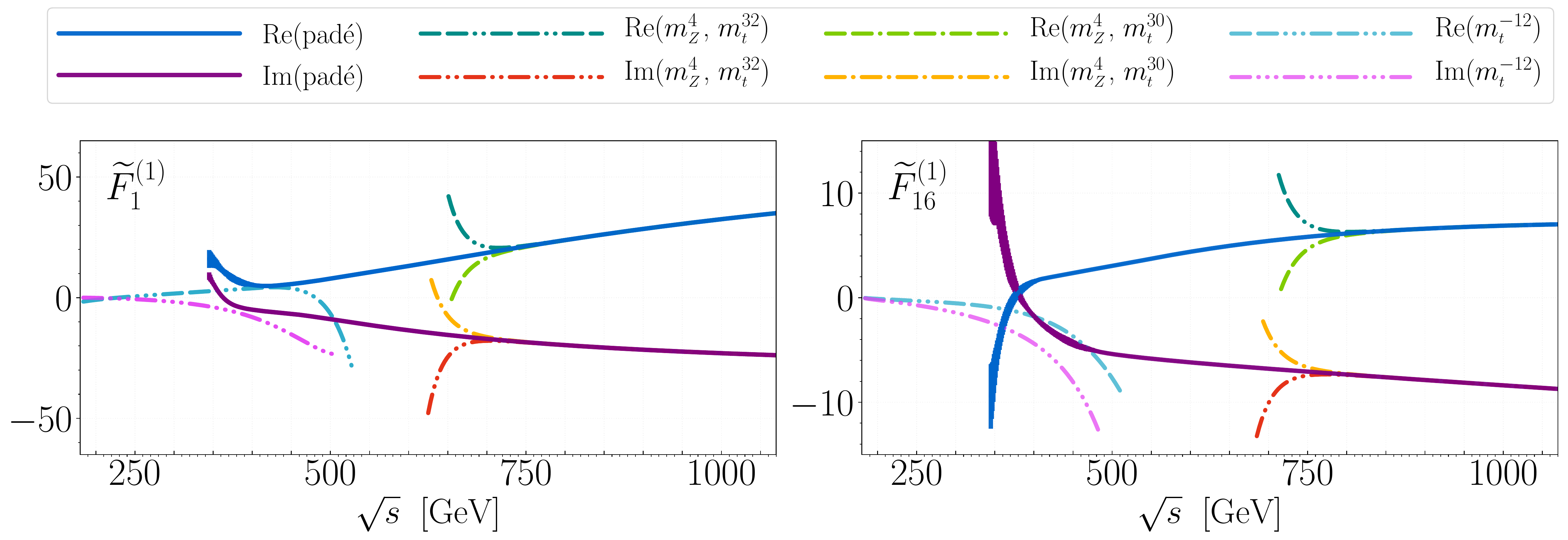}
  \caption{\label{fig::T_NLO}The NLO form factor of the tensor structures $T_1$ and
    $T_{16}$ as a function of $\sqrt{s}$, plotted for $\theta=\pi/2$.
    Both the real and imaginary parts are shown.
    Dash-dotted and dotted lines correspond to the 
    high-energy and large-$m_t$ results. The solid lines/bands represent the
    Pad\'e-improved predictions and their uncertainties.}
\end{figure}

In this Section we apply the approximation procedure of Section~\ref{sec::pade}
to our results for NLO quantities. We begin by considering two example form factors,
in the orthogonal basis of Section~\ref{sec::orthog-tensor-basis}: $\widetilde{F}_1^{(1)}$
and $\widetilde{F}_{16}^{(1)}$, renormalized and infrared-subtracted according to
Eq.~(\ref{eq::Ftil}).
In Fig.~\ref{fig::T_NLO} we show their real and imaginary parts as functions of $\sqrt{s}$.
Similar plots for the full set of 18 form factors can be found in
Appendix~\ref{app::T_NLO}.
The plots contain curves which show the large-$m_t$ expansion (to order $1/m_t^{12}$)
and the high-energy expansion (to order $m_t^{30} m_Z^4$ and $m_t^{32} m_Z^4$).
Just as at leading order, the high energy expansion does not converge below
$\sqrt{s}\approx 750$GeV. The solid curves show the Pad\'e-improved approximations
of both the real and imaginary parts of the form factors. In the case of
$\widetilde{F}_1^{(1)}$, the plot suggests that for real and imaginary parts the
curves merge smoothly into the large-$m_t$ expansion.
In the case of $\widetilde{F}_{16}^{(1)}$ we expect a resonance-like structure
(as for ${F}_{16}^{(0)}$ in Fig.~\ref{fig::T_LO})
which is also indicated by the Pad\'e curves.
The thickness of the solid curves reflect our estimate of the uncertainty due to
the approximation procedure, as defined in Eq.~(\ref{eq::alpha_delta_new}).

Although not the focus of this paper, the form factors also receive contributions
from massless quarks running in the loop. We extend the notation of
Section~\ref{sec::technical} and define
\begin{eqnarray}
  \overline{F}_i^{(0)} &=& \sum_{f=u,d,s,c,b} 
                     \left( {F}_{i,\rm ml,box}^{(0),v_f}
                     + {F}_{i,\rm ml,box}^{(0),a_f}
                     \right) + {F}_{i}^{(0)}
                     \,,
                     \label{eq::barF}
\end{eqnarray}
where for up-type quarks $v_f$ and $a_f$ are as given in Eq.~(\ref{eq::vtat}) and
for down-type quarks they are given by
\begin{align}
  v_b &= -\frac{1}{2}+\frac{2}{3}\sin^2\theta_W\,,\qquad
  a_b = -\frac{1}{2}
          \,.
          \label{eq::vbab}
\end{align}
In Eq.~(\ref{eq::barF}) ${F}_i^{(0)}$ corresponds to the LO top quark form factors
${f}_i^{(0)}$ of Eq.~(\ref{eq::F_tri_box}), in the orthogonal basis of
Section~\ref{sec::orthog-tensor-basis}. There is no contribution from
${F}_{i,\rm ml,tri}^{(0)}$ since it is
heavily suppressed by the Yukawa couplings of the light quarks and the Higgs boson.
We obtain ${F}_{i,\rm ml,box}^{(0),v_f}$ and ${F}_{i,\rm ml,box}^{(0),a_f}$ by
taking the massless limit of the exact LO expressions numerically.
Although the NLO massless contributions are known
(see Refs.~\cite{vonManteuffel:2015msa,Caola:2015ila}), they are not relevant for
our discussion of the quality of the approximations of the NLO top quark-loop
contributions. They can be added easily to the final result since they only
interfere with the exact LO expressions.

In terms of these $\overline{F}_i^{(0)}$, we now define the finite, virtual contribution to
the differential cross section (in analogy to $gg\to HH$~\cite{Heinrich:2017kxx,Davies:2019dfy}),
\begin{eqnarray}
  \widetilde{\mathcal{V}}_{\textnormal{fin}} &=&
  \frac{\alpha_s^2\left(\mu\right)}{\pi^2}\frac{G_F^2 m_Z^4}{32}
                                                 \sum_{i=1}^{18}
  \left[   C_i + 2\left( {\overline{F}_i^{(0)*}}
    \widetilde{F}_i^{(1)} + 
    \overline{F}_i^{(0)} {\widetilde{F}_i^{(1)*}}\right) \right] \,,
      \label{eq::Vtil}
\end{eqnarray}
where $C_i$ is defined by
\begin{eqnarray}
  C_i &=& \left|\overline{F}_i^{(0)}\right|^2 C_A
        \left(
        \pi^2 - \log^2\frac{\mu^2}{s}
        \right)
        \,.
          \label{eq::Ci}
\end{eqnarray}

In Eq.~(\ref{eq::Vtil}) $\alpha_s$ corresponds to the five-flavour strong
coupling constant.  We introduce the quantity
\begin{eqnarray}
  \mathcal{V}_{\textnormal{fin}} 
  &=& \frac{\widetilde{\mathcal{V}}_{\textnormal{fin}}}{\alpha_s^2(\mu)}
      \,,
\end{eqnarray}
which is discussed in the following. For the renormalization scale we choose
$\mu^2=s/4$.

When evaluating $\mathcal{V}_{\textnormal{fin}}$ we use exact expressions
for the LO form factors and our high-energy expansions for the
NLO parts. Exact results are known for the two-loop triangle
form factors~\cite{Harlander:2005rq,Anastasiou:2006hc,Aglietti:2006tp},
however as shown in~\cite{Davies:2018qvx}, the high-energy expansions
reproduce the exact result almost down to the top threshold, which justifies
our use of the expansions to evaluate these contributions also.

\begin{figure}[t]
\begin{center}
  \begin{tabular}{cc}
	\hspace{-4mm}
    \includegraphics[width=0.49\textwidth]{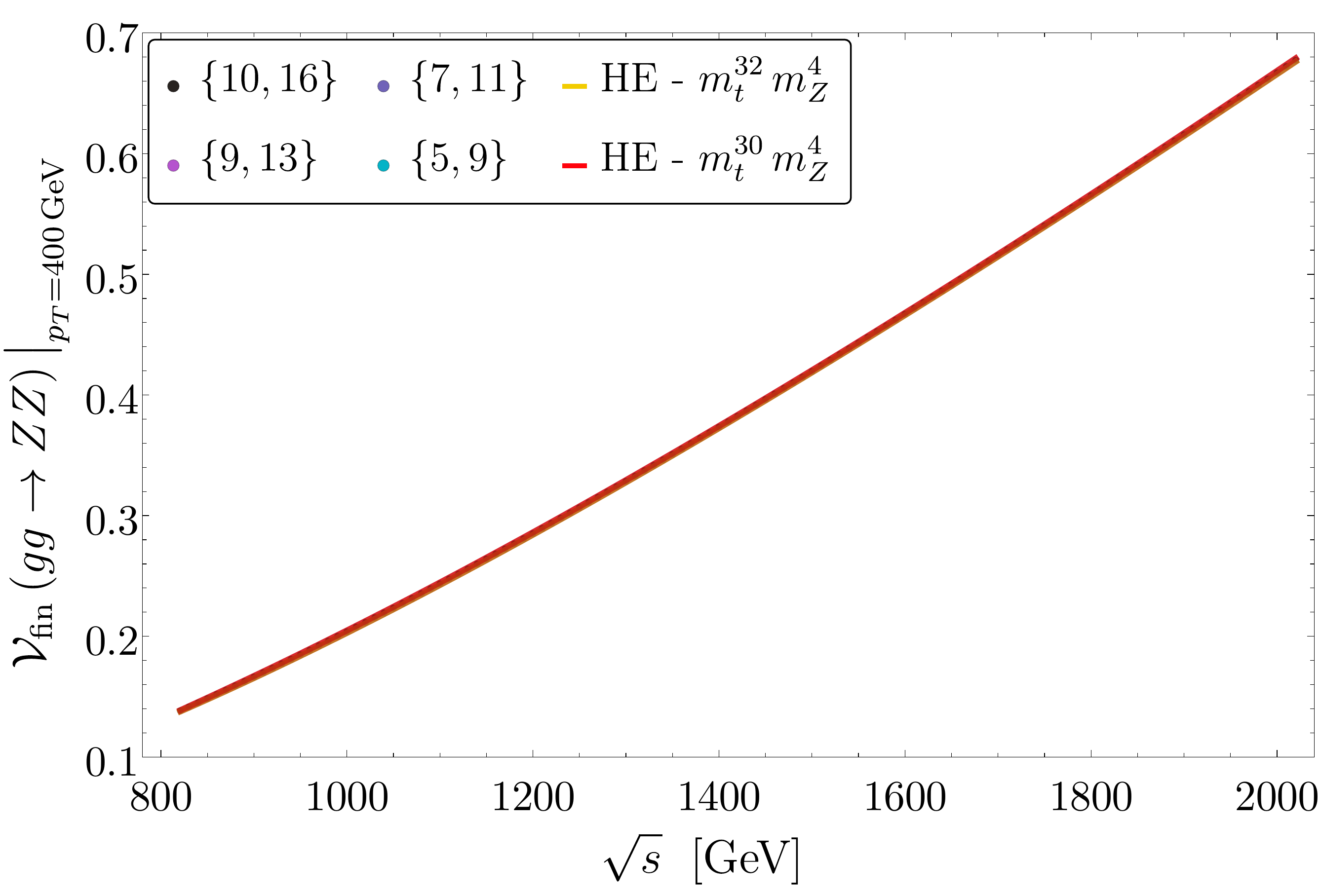} &
    \includegraphics[width=0.49\textwidth]{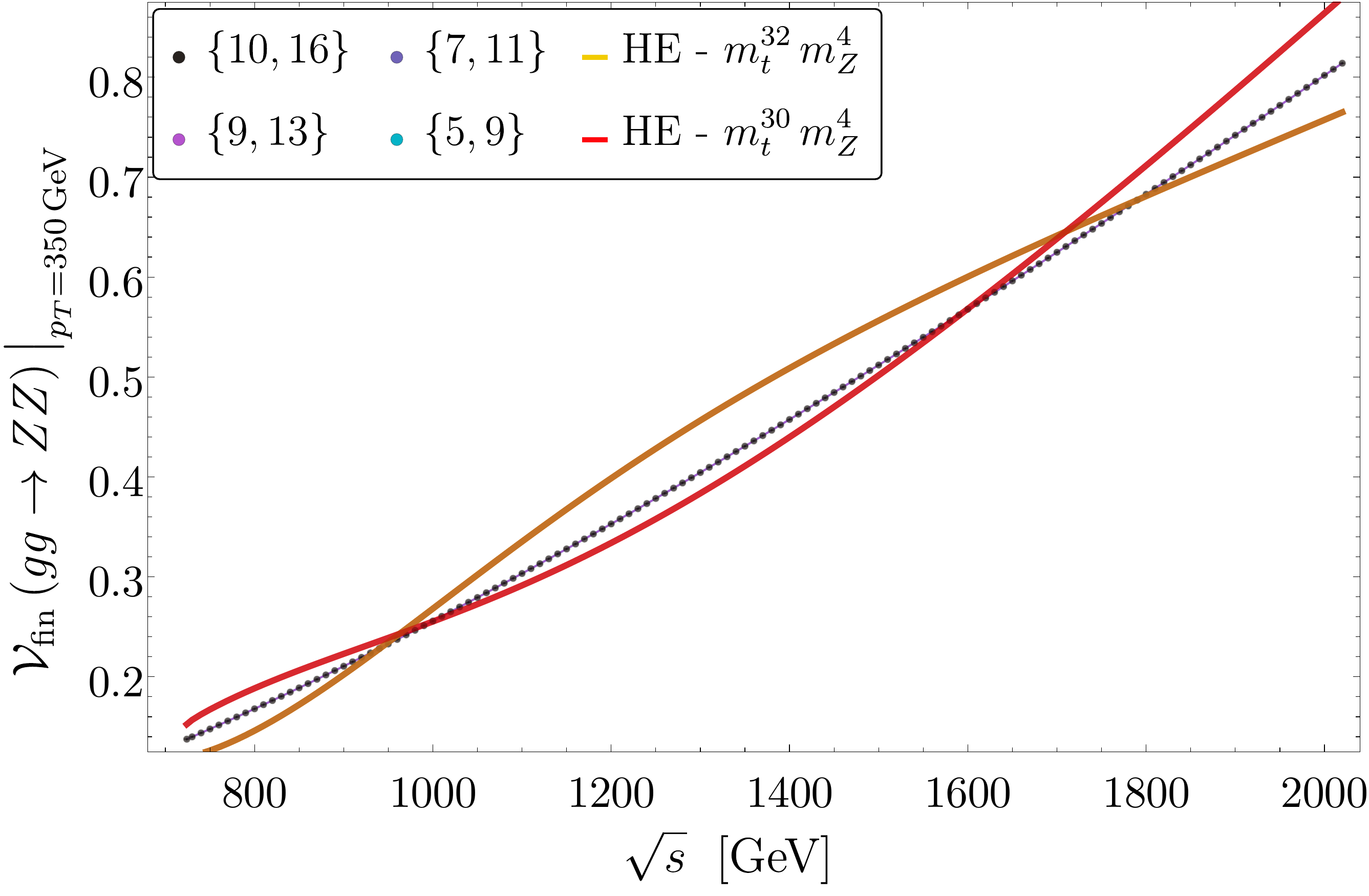} \\
  \end{tabular}
  \caption{\label{fig::Vfin_1}${\cal V}_{\rm fin}$ as a function of $\sqrt{s}$
    for $p_T$ values of 400 and 350~GeV. The curves denoted ``HE'' show
    the high-energy expansion to orders $m_t^{30}m_Z^4$ and $m_t^{32}m_Z^4$.}
\end{center}
\end{figure}

\begin{figure}[t]
  \begin{tabular}{cc}
	\hspace{-4mm}
    \includegraphics[width=0.49\textwidth]{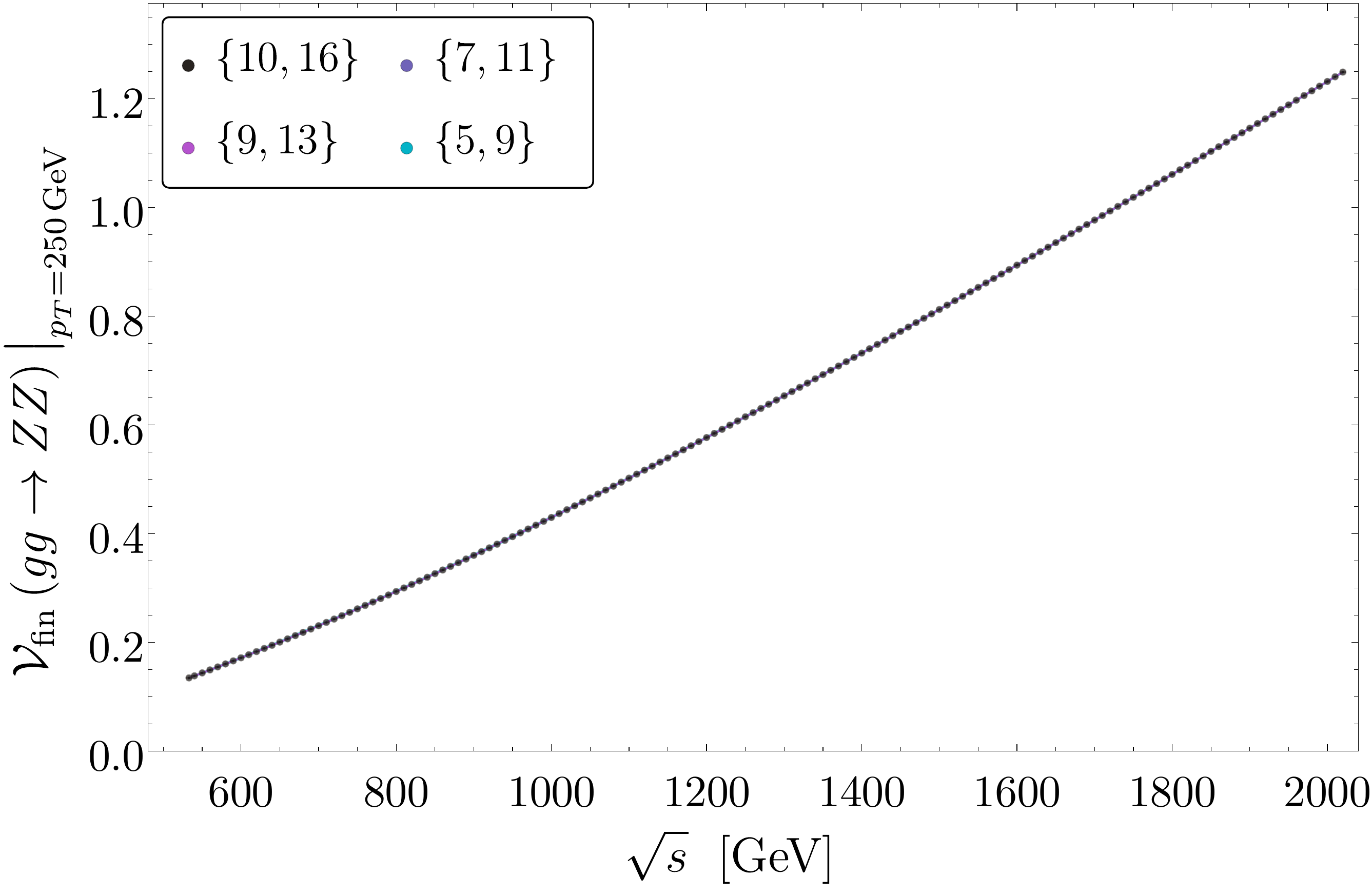} &
    \includegraphics[width=0.49\textwidth]{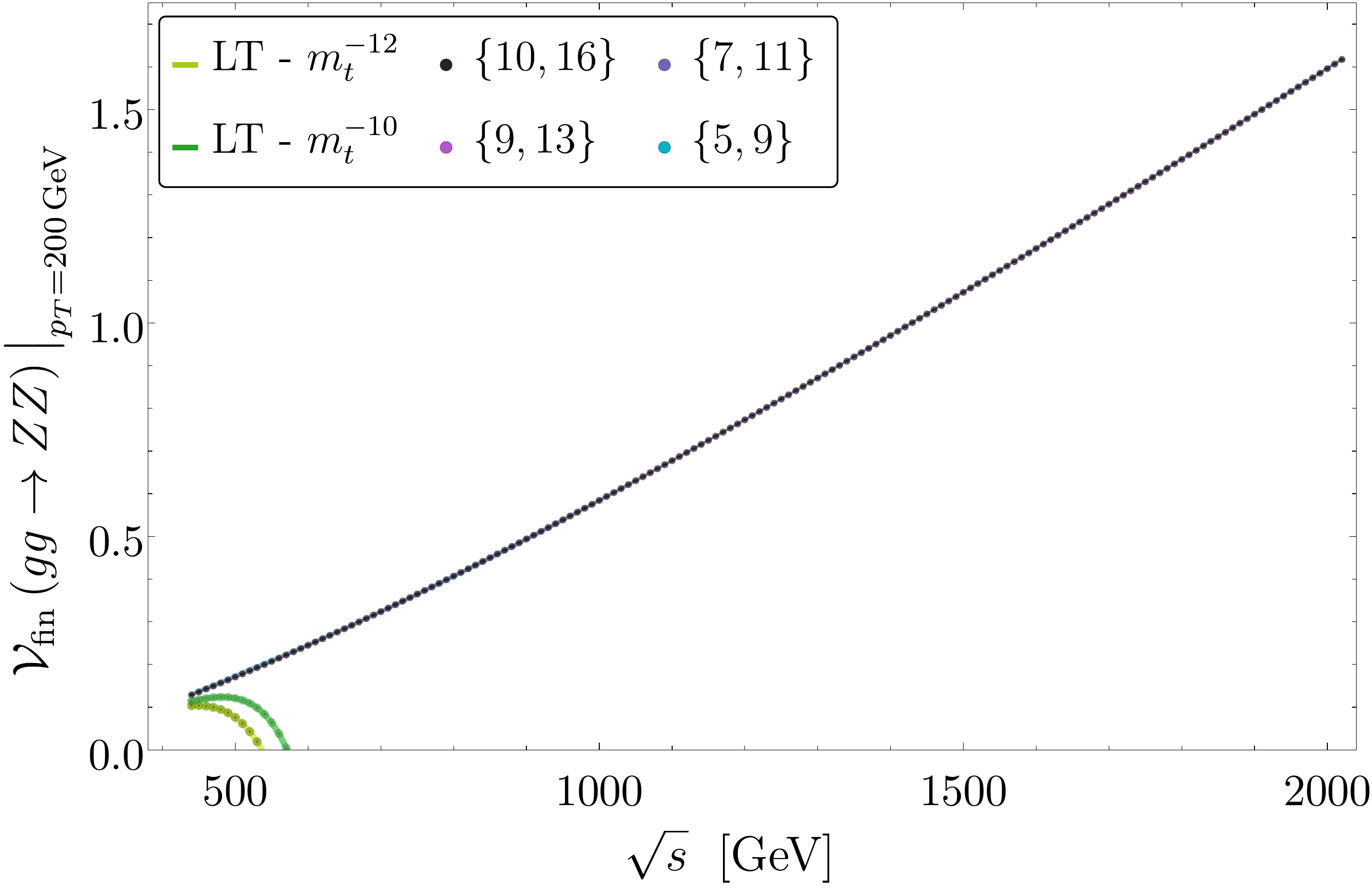} \\
	\hspace{-4mm}
    \includegraphics[width=0.49\textwidth]{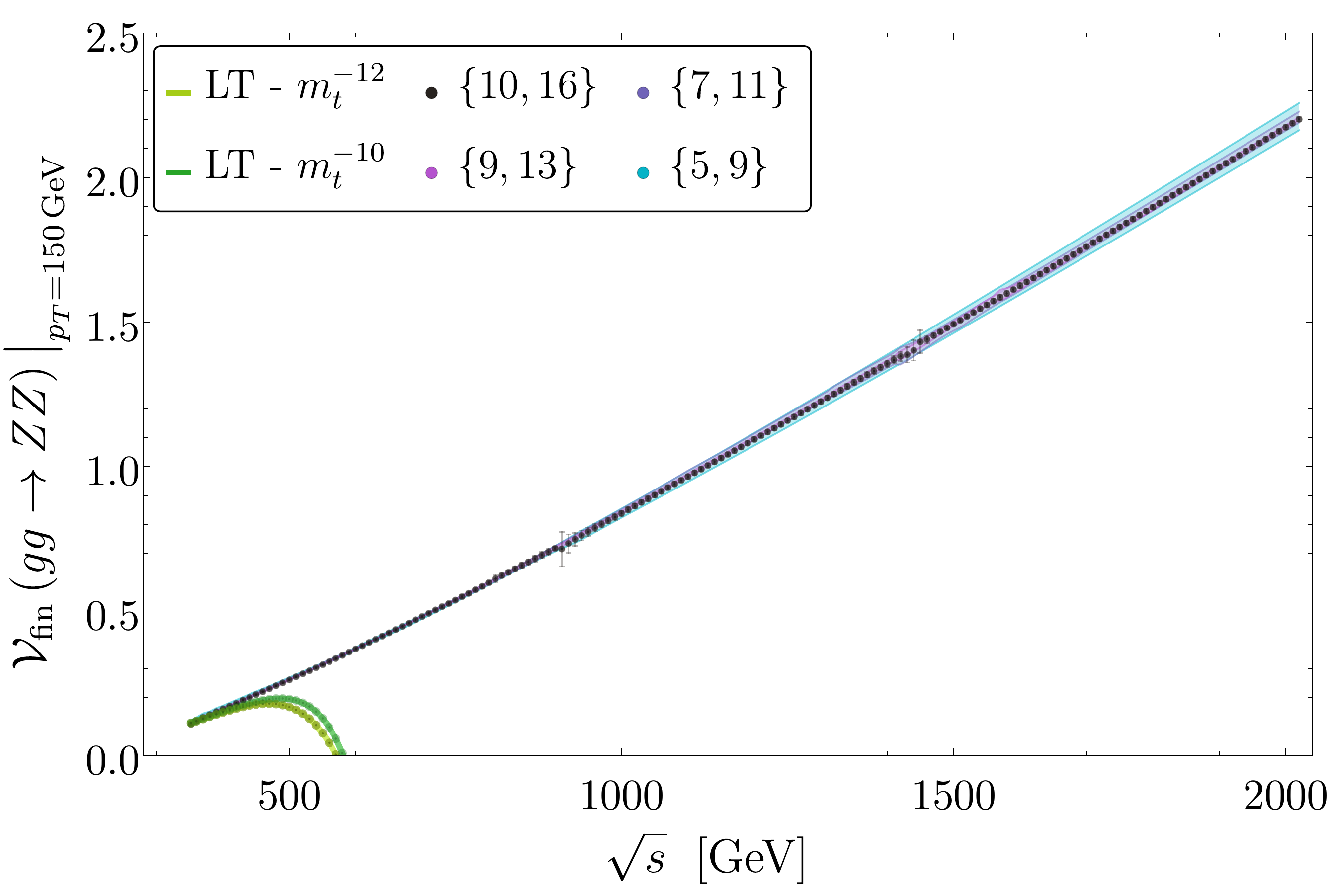} &
    \includegraphics[width=0.49\textwidth]{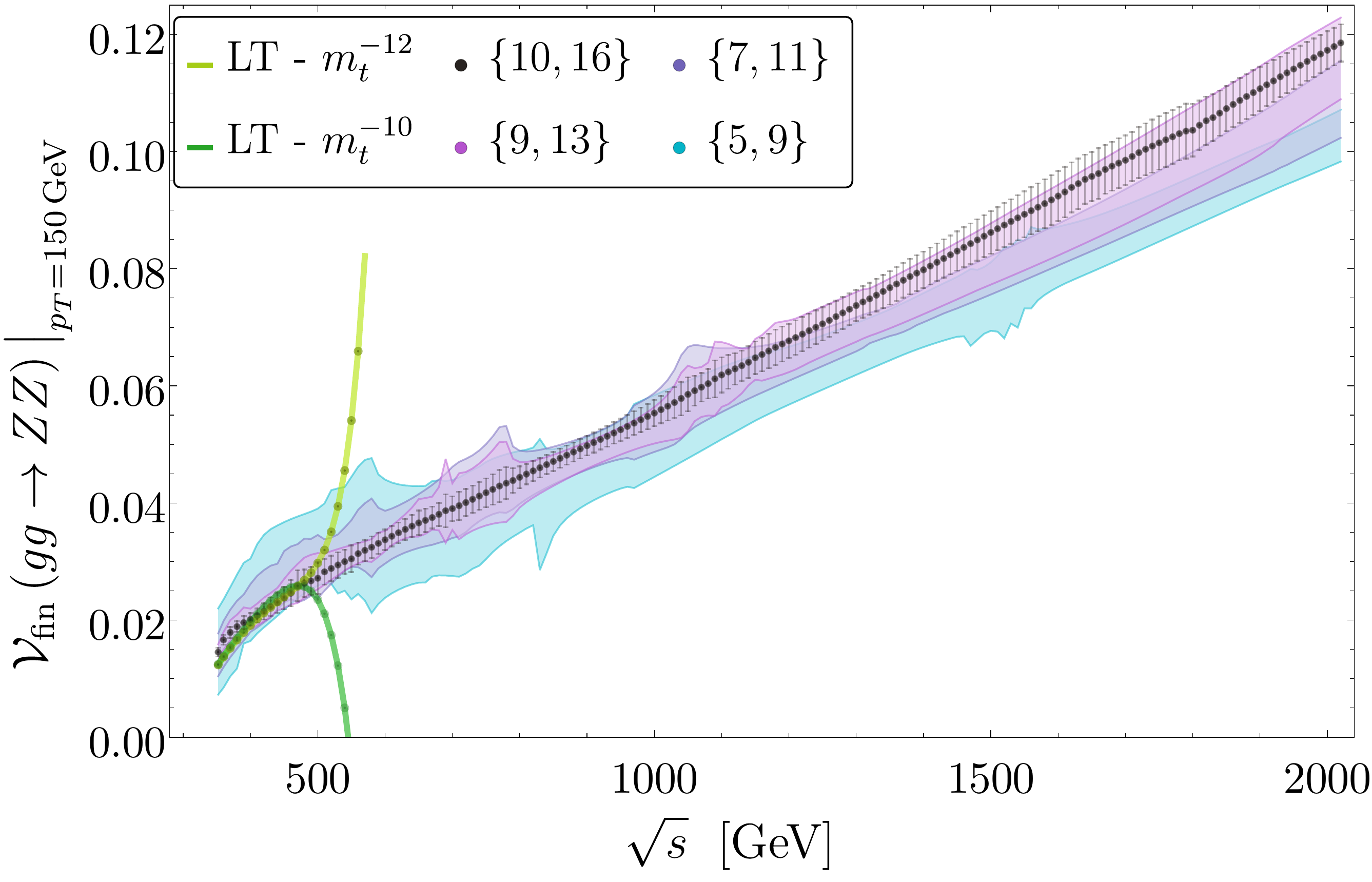} \\
  \end{tabular}
  \caption{\label{fig::Vfin_2}${\cal V}_{\rm fin}$ as a function of $\sqrt{s}$
  for $p_T$ values of 250, 200 and 150~GeV. The curves denoted ``LT'' show the
  large-$m_t$ expansion to orders $m_t^{-10}$ and $m_t^{-12}$. The bottom-right
  plot excludes the light quark contributions introduced in Eq.~(\ref{eq::barF}).}
\end{figure}

In Figs.~\ref{fig::Vfin_1} and~\ref{fig::Vfin_2} we show ${\cal V}_{\rm fin}$
as a function of $\sqrt{s}$ for different values of $p_T$.
In all cases we show Pad\'{e}-improved results
which are obtained by applying the method from Section~\ref{sec::pade} to
${\cal V}_{\rm fin}$. We start with the numerical evaluation of the form
factors $\widetilde{f}_i$, keeping the variable $x$ as introduced in
Section~\ref{sec::pade}, perform the basis change to $\widetilde{F}_i$ and then construct
${\cal V}_{\rm fin}$ as an expansion in $x$. At this point we apply the
procedure outlined in Section~\ref{sec::pade}.
We take into account the different sets of Pad\'e approximants listed below
Eq.~(\ref{eq::N_low_high}). Our best prediction, \{10,16\}, is shown as black points.
For higher values of $p_T$ (Fig.~\ref{fig::Vfin_1}) we show in addition
two curves from the high-energy expansion.  For $p_T=400$~GeV they lie on top of
each other and agree with the Pad\'e predictions.
For $p_T=350$~GeV the two high-energy expansion curves differ from
each other.
The Pad\'e approximations show a stable behaviour, demonstrated by
the fact that even with little input (\{5,9\}) no significant
uncertainty is observed.

For lower values of $p_T$ (Fig.~\ref{fig::Vfin_2})
the high-energy expansion curves lie completely outside of the range of the
plot axes. Nonetheless stable Pad\'e predictions are observed, even for the
low $p_T$ value of 150~GeV. For this value one observes that the higher orders
in the $m_t$ expansion are crucial to obtain estimates with small uncertainties.
The bottom-right panel of Fig.~\ref{fig::Vfin_2} shows the same approximations,
but excludes the light quark contributions introduced in Eq.~(\ref{eq::barF}).
This shows in more detail the improvement of the \{10,16\} approximation
with respect to, for example, the \{5,9\} approximation.

The results presented in Figs.~\ref{fig::Vfin_1} and~\ref{fig::Vfin_2}
make predictions for ${\cal V}_{\rm fin}$ which should eventually be
confronted with numerical results as, e.g., announced in Ref.~\cite{Agarwal:2019rag}.
To this end we provide, in the ancillary files of this paper~\cite{progdata},
a simple \texttt{C++} program which interpolates a pre-evaluated grid of
${\cal V}_{\rm fin}$ points and uncertainties, evaluated with \{10,16\}
for $\mu^2 = s/4$, using routines from the GNU Scientific Library~\cite{GSL}.
It can thus be used to reproduce the black points and uncertainty bars
of Figs.~\ref{fig::Vfin_1} and \ref{fig::Vfin_2}.

%- }}}
%- {{{ Conclusions:

\section{Conclusions}
\label{sec::concl}
We compute NLO QCD corrections to the process $gg\to ZZ$
induced by virtual top quark loops. We concentrate on the
high-energy limit which corresponds to an expansion
in the parameters $m_t^2/s$, $m_t^2/t$ and $m_t^2/u$.
We furthermore expand for small $Z$ boson masses and show, at LO,
that three expansion terms are sufficient to obtain per-mille accuracy.
Analytic results, including terms up to order $m_t^{32} m_Z^4$,
are presented for all 20 form factors in the ancillary file~\cite{progdata}.
Additionally we include in this file the large-$m_t$ expansions of these 20
form factors, up to order $1/m_t^{12}$.

Using simple tensor structures as a starting point, we construct
an orthogonal basis which is convenient when
computing the squared amplitude. Alternatively we also provide
LO and NLO results for the helicity amplitudes.

We extend the radius of convergence of the high-energy expansions
with the help of Pad\'e approximations. Our method provides both
a central value and an uncertainty estimate. This
is validated by comparisons to known exact
results at LO. The Pad\'e method is applied both to the form factors
and the NLO virtual corrections to the differential
cross section. In the latter case we include in our
predictions also the LO contributions which originate
from massless quark loops.
In this setup the interference of the
one- and two-loop top quark contributions amounts to
about 5\% as can be seen from Fig.~\ref{fig::Vfin_2}.

%- }}}

%- {{{ Ackn.:
 
\section*{Acknowledgements}

This work is supported by the BMBF grant 05H15VKCCA.
DW acknowledges the support of the DFG-funded Doctoral School KSETA.

%- }}}

%- {{{ appendix:

\begin{appendix}

%- {{{ Basis change:

\section{\label{app::T}Basis change}

In this Appendix we provide the basis change relations from the simple tensor
structures $S_i^{\mu\nu\rho\sigma}$ of Eq.~(\ref{eq::calT}) to the orthogonal tensors 
$T_i^{\mu\nu\rho\sigma}$ of Section~\ref{sec::orthog-tensor-basis}. They are given by
\begin{footnotesize}
\begin{alignat}{4}
&T_{1}^{\mu\nu\rho\sigma}&&=\;\;&& 
S_{17}^{\mu\nu\rho\sigma}  \frac{m_Z^2}{p_T^2 (p_1\!\cdot\!p_3) (p_2\!\cdot\!p_3)}\:,
 &&\label{app:ggzz:basischange:eq:t1}\\ 
&T_{2}^{\mu\nu\rho\sigma}&&=\;\;&& 
S_{17}^{\mu\nu\rho\sigma}  \frac{\left(m_Z^2-p_T^2\right)^2}{4 p_T^4 (p_1\!\cdot\!p_3) (p_2\!\cdot\!p_3)} 
  +
S_{18}^{\mu\nu\rho\sigma}  \frac{\left(m_Z^2-p_T^2\right) (p_2\!\cdot\!p_3)}{2 p_T^4 (p_1\!\cdot\!p_2) (p_1\!\cdot\!p_3)} 
  +
S_{19}^{\mu\nu\rho\sigma}  \frac{\left(m_Z^2-p_T^2\right) (p_1\!\cdot\!p_3)}{2 p_T^4 (p_1\!\cdot\!p_2) (p_2\!\cdot\!p_3)} 
  &&\\\nonumber  & && \;+&&
S_{20}^{\mu\nu\rho\sigma}  \frac{(p_1\!\cdot\!p_3) (p_2\!\cdot\!p_3)}{p_T^4 (p_1\!\cdot\!p_2)^2}\:,
 &&\\ 
&T_{3}^{\mu\nu\rho\sigma}&&=\;\;&& 
S_{9}^{\mu\nu\rho\sigma}  \frac{m_Z^2}{(p_1\!\cdot\!p_3) (p_2\!\cdot\!p_3)}
  +
S_{17}^{\mu\nu\rho\sigma}  \frac{m_Z^2}{p_T^2 (p_1\!\cdot\!p_3) (p_2\!\cdot\!p_3)} \:,
 &&\\ 
&T_{4}^{\mu\nu\rho\sigma}&&=\;\;&& 
S_{9}^{\mu\nu\rho\sigma}  \frac{\left(m_Z^2-p_T^2\right)^2}{4 p_T^2 (p_1\!\cdot\!p_3) (p_2\!\cdot\!p_3)}  
  +
S_{10}^{\mu\nu\rho\sigma}  \frac{\left(m_Z^2-p_T^2\right) (p_2\!\cdot\!p_3)}{2 p_T^2 (p_1\!\cdot\!p_2) (p_1\!\cdot\!p_3)} 
  +
S_{11}^{\mu\nu\rho\sigma}  \frac{\left(m_Z^2-p_T^2\right) (p_1\!\cdot\!p_3)}{2 p_T^2 (p_1\!\cdot\!p_2) (p_2\!\cdot\!p_3)}
  &&\\\nonumber  & && \;+&&
S_{12}^{\mu\nu\rho\sigma}  \frac{(p_1\!\cdot\!p_3) (p_2\!\cdot\!p_3)}{p_T^2 (p_1\!\cdot\!p_2)^2}
  +
S_{17}^{\mu\nu\rho\sigma}  \frac{\left(m_Z^2-p_T^2\right)^2}{4 p_T^4 (p_1\!\cdot\!p_3) (p_2\!\cdot\!p_3)} 
  +
S_{18}^{\mu\nu\rho\sigma}  \frac{\left(m_Z^2-p_T^2\right) (p_2\!\cdot\!p_3)}{2 p_T^4 (p_1\!\cdot\!p_2) (p_1\!\cdot\!p_3)} 
  &&\\\nonumber  & && \;+&&
S_{19}^{\mu\nu\rho\sigma}  \frac{\left(m_Z^2-p_T^2\right) (p_1\!\cdot\!p_3)}{2 p_T^4 (p_1\!\cdot\!p_2) (p_2\!\cdot\!p_3)} 
  +
S_{20}^{\mu\nu\rho\sigma}  \frac{(p_1\!\cdot\!p_3) (p_2\!\cdot\!p_3)}{p_T^4 (p_1\!\cdot\!p_2)^2} \:,
 &&\\ 
&T_{5}^{\mu\nu\rho\sigma}&&=\;\;&& 
S_{4}^{\mu\nu\rho\sigma}  \frac{\left(m_Z^2-p_T^2\right)}{2 p_T^2 (p_1\!\cdot\!p_3)} 
  +
S_{5}^{\mu\nu\rho\sigma}  \frac{(p_1\!\cdot\!p_3)}{p_T^2 (p_1\!\cdot\!p_2)}
  +
S_{17}^{\mu\nu\rho\sigma}  \frac{\left(m_Z^2-p_T^2\right)}{2 p_T^4 (p_1\!\cdot\!p_2)} 
  +
S_{18}^{\mu\nu\rho\sigma}  \frac{\left(m_Z^2-p_T^2\right) (p_2\!\cdot\!p_3)}{2 p_T^4 (p_1\!\cdot\!p_2) (p_1\!\cdot\!p_3)} 
  &&\\\nonumber  & && \;+&&
S_{19}^{\mu\nu\rho\sigma}  \frac{(p_1\!\cdot\!p_3)^2}{p_T^4 (p_1\!\cdot\!p_2)^2} 
  +
S_{20}^{\mu\nu\rho\sigma}  \frac{(p_1\!\cdot\!p_3) (p_2\!\cdot\!p_3)}{p_T^4 (p_1\!\cdot\!p_2)^2} \:,
 &&\\ 
&T_{6}^{\mu\nu\rho\sigma}&&=\;\;&& 
S_{13}^{\mu\nu\rho\sigma}  \frac{\left(m_Z^2-p_T^2\right)}{2 p_T^2 (p_2\!\cdot\!p_3)} 
  +
S_{14}^{\mu\nu\rho\sigma}  \frac{(p_2\!\cdot\!p_3)}{p_T^2 (p_1\!\cdot\!p_2)}
  -
S_{17}^{\mu\nu\rho\sigma} \frac{\left(m_Z^2-p_T^2\right)}{2 p_T^4 (p_1\!\cdot\!p_2)} 
  -
S_{18}^{\mu\nu\rho\sigma} \frac{(p_2\!\cdot\!p_3)^2}{p_T^4 (p_1\!\cdot\!p_2)^2} 
  &&\\\nonumber  & && \;-&&
S_{19}^{\mu\nu\rho\sigma} \frac{\left(m_Z^2-p_T^2\right) (p_1\!\cdot\!p_3)}{2 p_T^4 (p_1\!\cdot\!p_2) (p_2\!\cdot\!p_3)} 
  -
S_{20}^{\mu\nu\rho\sigma} \frac{(p_1\!\cdot\!p_3) (p_2\!\cdot\!p_3)}{p_T^4 (p_1\!\cdot\!p_2)^2} \:,
 &&\\ 
&T_{7}^{\mu\nu\rho\sigma}&&=\;\;&& 
S_{6}^{\mu\nu\rho\sigma}  \frac{(p_2\!\cdot\!p_3)}{p_T^2 (p_1\!\cdot\!p_2)}
  +
S_{7}^{\mu\nu\rho\sigma}  \frac{\left(m_Z^2-p_T^2\right)}{2 p_T^2 (p_2\!\cdot\!p_3)} 
  +
S_{17}^{\mu\nu\rho\sigma}  \frac{(p_1\!\cdot\!p_3) (p_2\!\cdot\!p_3)}{p_T^4 (p_1\!\cdot\!p_2)^2} 
  +
S_{18}^{\mu\nu\rho\sigma}  \frac{(p_2\!\cdot\!p_3)^2}{p_T^4 (p_1\!\cdot\!p_2)^2} 
  &&\\\nonumber  & && \;+&&
S_{19}^{\mu\nu\rho\sigma}  \frac{\left(m_Z^2-p_T^2\right) (p_1\!\cdot\!p_3)}{2 p_T^4 (p_1\!\cdot\!p_2) (p_2\!\cdot\!p_3)} 
  +
S_{20}^{\mu\nu\rho\sigma}  \frac{\left(m_Z^2-p_T^2\right)}{2 p_T^4 (p_1\!\cdot\!p_2)} \:,
 &&\\
&T_{8}^{\mu\nu\rho\sigma}&&=\;\;&& 
S_{15}^{\mu\nu\rho\sigma} \frac{(p_1\!\cdot\!p_3)}{p_T^2 (p_1\!\cdot\!p_2)}
  +
S_{16}^{\mu\nu\rho\sigma} \frac{\left(m_Z^2-p_T^2\right)}{2 p_T^2 (p_1\!\cdot\!p_3)} 
  +
S_{17}^{\mu\nu\rho\sigma} \frac{(p_1\!\cdot\!p_3) (p_2\!\cdot\!p_3)}{p_T^4 (p_1\!\cdot\!p_2)^2} 
  -
S_{18}^{\mu\nu\rho\sigma} \frac{\left(m_Z^2-p_T^2\right) (p_2\!\cdot\!p_3)}{2 p_T^4 (p_1\!\cdot\!p_2) (p_1\!\cdot\!p_3)} 
  &&\\\nonumber  & && \;-&&
S_{19}^{\mu\nu\rho\sigma} \frac{(p_1\!\cdot\!p_3)^2}{p_T^4 (p_1\!\cdot\!p_2)^2} 
  -
S_{20}^{\mu\nu\rho\sigma} \frac{\left(m_Z^2-p_T^2\right)}{2 p_T^4 (p_1\!\cdot\!p_2)} \:,
 &&
\\
%\end{alignat}
%
%\begin{alignat}{4}
%%
%%
&T_{9}^{\mu\nu\rho\sigma}&&=\;\;&& 
S_{8}^{\mu\nu\rho\sigma}  \frac{1}{p_T^2}
  -
S_{17}^{\mu\nu\rho\sigma}  \frac{(p_1\!\cdot\!p_3) (p_2\!\cdot\!p_3)}{p_T^4 (p_1\!\cdot\!p_2)^2} 
  -
S_{18}^{\mu\nu\rho\sigma} \frac{p_T^2 (p_1\!\cdot\!p_2)+(p_2\!\cdot\!p_3)^2}{p_T^4 (p_1\!\cdot\!p_2)^2} 
  &&\\\nonumber  & && \;-&&
S_{19}^{\mu\nu\rho\sigma} \frac{p_T^2 (p_1\!\cdot\!p_2)+(p_1\!\cdot\!p_3)^2}{p_T^4 (p_1\!\cdot\!p_2)^2} 
  -
S_{20}^{\mu\nu\rho\sigma} \frac{(p_1\!\cdot\!p_3) (p_2\!\cdot\!p_3)}{p_T^4 (p_1\!\cdot\!p_2)^2} \:,
 &&\\
&T_{10}^{\mu\nu\rho\sigma}&&=\;\;&& 
S_{3}^{\mu\nu\rho\sigma}
  -
S_{4}^{\mu\nu\rho\sigma} \frac{(p_2\!\cdot\!p_3)}{p_T^2 (p_1\!\cdot\!p_2)} 
  -
S_{5}^{\mu\nu\rho\sigma} \frac{(p_1\!\cdot\!p_3)}{p_T^2 (p_1\!\cdot\!p_2)} 
  +
S_{15}^{\mu\nu\rho\sigma}  \frac{(p_1\!\cdot\!p_3)}{p_T^2 (p_1\!\cdot\!p_2)} 
  +
S_{16}^{\mu\nu\rho\sigma}  \frac{(p_2\!\cdot\!p_3)}{p_T^2 (p_1\!\cdot\!p_2)} 
  &&\\\nonumber  & && \;-&&
S_{17}^{\mu\nu\rho\sigma}  \frac{(p_1\!\cdot\!p_3) (p_2\!\cdot\!p_3)}{p_T^4 (p_1\!\cdot\!p_2)^2} 
  -
S_{18}^{\mu\nu\rho\sigma} \frac{(p_2\!\cdot\!p_3)^2}{p_T^4 (p_1\!\cdot\!p_2)^2} 
  -
S_{19}^{\mu\nu\rho\sigma} \frac{(p_1\!\cdot\!p_3)^2}{p_T^4 (p_1\!\cdot\!p_2)^2} 
  -
S_{20}^{\mu\nu\rho\sigma} \frac{(p_1\!\cdot\!p_3) (p_2\!\cdot\!p_3)}{p_T^4 (p_1\!\cdot\!p_2)^2} \:,
 &&\\
&T_{11}^{\mu\nu\rho\sigma}&&=\;\;&& 
S_{17}^{\mu\nu\rho\sigma}  \frac{m_Z^2 \left(m_Z^2-p_T^2\right)}{2 p_T^2 (p_1\!\cdot\!p_3) (p_2\!\cdot\!p_3)} 
  +
S_{19}^{\mu\nu\rho\sigma}  \frac{m_Z^2 (p_1\!\cdot\!p_3)}{p_T^2 (p_1\!\cdot\!p_2) (p_2\!\cdot\!p_3)}\:,
 &&\\ 
&T_{12}^{\mu\nu\rho\sigma}&&=\;\;&& 
S_{17}^{\mu\nu\rho\sigma}  \frac{m_Z^2 \left(m_Z^2-p_T^2\right)}{2 p_T^2 (p_1\!\cdot\!p_3) (p_2\!\cdot\!p_3)} 
  +
S_{18}^{\mu\nu\rho\sigma}  \frac{m_Z^2 (p_2\!\cdot\!p_3)}{p_T^2 (p_1\!\cdot\!p_2) (p_1\!\cdot\!p_3)}\:,
 &&\\ 
&T_{13}^{\mu\nu\rho\sigma}&&=\;\;&& 
S_{7}^{\mu\nu\rho\sigma}  \frac{m_Z^2}{(p_2\!\cdot\!p_3)}
  +
S_{19}^{\mu\nu\rho\sigma}  \frac{m_Z^2 (p_1\!\cdot\!p_3)}{p_T^2 (p_1\!\cdot\!p_2) (p_2\!\cdot\!p_3)} 
  +
S_{20}^{\mu\nu\rho\sigma}  \frac{m_Z^2}{p_T^2 (p_1\!\cdot\!p_2)} \:,
 &&\\ 
&T_{14}^{\mu\nu\rho\sigma}&&=\;\;&& 
S_{16}^{\mu\nu\rho\sigma}  \frac{m_Z^2}{(p_1\!\cdot\!p_3)}
  -
% - sign between S18 and coefficient removed here
S_{18}^{\mu\nu\rho\sigma} \frac{m_Z^2 (p_2\!\cdot\!p_3)}{p_T^2 (p_1\!\cdot\!p_2) (p_1\!\cdot\!p_3)} 
  -
S_{20}^{\mu\nu\rho\sigma} \frac{m_Z^2}{p_T^2 (p_1\!\cdot\!p_2)} \:,
 &&\\ 
&T_{15}^{\mu\nu\rho\sigma}&&=\;\;&& 
  +
S_{4}^{\mu\nu\rho\sigma}  \frac{m_Z^2}{(p_1\!\cdot\!p_3)}
  +
S_{18}^{\mu\nu\rho\sigma}  \frac{m_Z^2 (p_2\!\cdot\!p_3)}{p_T^2 (p_1\!\cdot\!p_2) (p_1\!\cdot\!p_3)} 
  +
S_{17}^{\mu\nu\rho\sigma}  \frac{m_Z^2}{p_T^2 (p_1\!\cdot\!p_2)} \:,
 &&\\ 
&T_{16}^{\mu\nu\rho\sigma}&&=\;\;&& 
S_{13}^{\mu\nu\rho\sigma}  \frac{m_Z^2}{(p_2\!\cdot\!p_3)}
  -
S_{17}^{\mu\nu\rho\sigma} \frac{m_Z^2}{p_T^2 (p_1\!\cdot\!p_2)} 
  -
S_{19}^{\mu\nu\rho\sigma}  \frac{m_Z^2 (p_1\!\cdot\!p_3)}{p_T^2 (p_1\!\cdot\!p_2) (p_2\!\cdot\!p_3)} \:,
 &&\\ 
&T_{17}^{\mu\nu\rho\sigma}&&=\;\;&& 
S_{9}^{\mu\nu\rho\sigma}  \frac{m_Z^2 \left(m_Z^2-p_T^2\right)}{2 (p_1\!\cdot\!p_3) (p_2\!\cdot\!p_3)} 
  +
S_{11}^{\mu\nu\rho\sigma}  \frac{m_Z^2 (p_1\!\cdot\!p_3)}{(p_1\!\cdot\!p_2) (p_2\!\cdot\!p_3)}
  +
S_{17}^{\mu\nu\rho\sigma}  \frac{m_Z^2 \left(m_Z^2-p_T^2\right)}{2 p_T^2 (p_1\!\cdot\!p_3) (p_2\!\cdot\!p_3)} 
  &&\\\nonumber  & && \;+&&
S_{19}^{\mu\nu\rho\sigma}  \frac{m_Z^2 (p_1\!\cdot\!p_3)}{p_T^2 (p_1\!\cdot\!p_2) (p_2\!\cdot\!p_3)} \:,
 &&\\ 
&T_{18}^{\mu\nu\rho\sigma}&&=\;\;&& 
S_{9}^{\mu\nu\rho\sigma}  \frac{m_Z^2 \left(m_Z^2-p_T^2\right)}{2 (p_1\!\cdot\!p_3) (p_2\!\cdot\!p_3)} 
  +
S_{10}^{\mu\nu\rho\sigma}  \frac{m_Z^2 (p_2\!\cdot\!p_3)}{(p_1\!\cdot\!p_2) (p_1\!\cdot\!p_3)}
  +
S_{17}^{\mu\nu\rho\sigma}  \frac{m_Z^2 \left(m_Z^2-p_T^2\right)}{2 p_T^2 (p_1\!\cdot\!p_3) (p_2\!\cdot\!p_3)} 
  &&\\\nonumber  & && \;+&&
S_{18}^{\mu\nu\rho\sigma}  \frac{m_Z^2 (p_2\!\cdot\!p_3)}{p_T^2 (p_1\!\cdot\!p_2) (p_1\!\cdot\!p_3)} \:,
 &&\\ 
&T_{19}^{\mu\nu\rho\sigma}&&=\;\;&& 
S_{2}^{\mu\nu\rho\sigma} 
  -
S_{3}^{\mu\nu\rho\sigma}
  +
S_{4}^{\mu\nu\rho\sigma}  \frac{(p_2\!\cdot\!p_3)}{p_T^2 (p_1\!\cdot\!p_2)} 
  +
S_{5}^{\mu\nu\rho\sigma}  \frac{(p_1\!\cdot\!p_3)}{p_T^2 (p_1\!\cdot\!p_2)} 
  -
S_{6}^{\mu\nu\rho\sigma} \frac{(p_2\!\cdot\!p_3)}{p_T^2 (p_1\!\cdot\!p_2)} 
  &&\\\nonumber  & && \;-&&
S_{7}^{\mu\nu\rho\sigma} \frac{(p_1\!\cdot\!p_3)}{p_T^2 (p_1\!\cdot\!p_2)} 
  +
S_{13}^{\mu\nu\rho\sigma}  \frac{(p_1\!\cdot\!p_3)}{p_T^2 (p_1\!\cdot\!p_2)} 
  +
S_{14}^{\mu\nu\rho\sigma}  \frac{(p_2\!\cdot\!p_3)}{p_T^2 (p_1\!\cdot\!p_2)} 
  -
S_{15}^{\mu\nu\rho\sigma} \frac{(p_1\!\cdot\!p_3)}{p_T^2 (p_1\!\cdot\!p_2)} 
  &&\\\nonumber  & && \;-&&
S_{16}^{\mu\nu\rho\sigma} \frac{(p_2\!\cdot\!p_3)}{p_T^2 (p_1\!\cdot\!p_2)} \:,
 &&\\
\label{app:ggzz:basischange:eq:t20}&T_{20}^{\mu\nu\rho\sigma}&&=\;\;&& 
S_{1}^{\mu\nu\rho\sigma} 
  -
S_{3}^{\mu\nu\rho\sigma}
  +
S_{4}^{\mu\nu\rho\sigma}  \frac{(p_2\!\cdot\!p_3)}{p_T^2 (p_1\!\cdot\!p_2)} 
  +
S_{5}^{\mu\nu\rho\sigma}  \frac{(p_1\!\cdot\!p_3)}{p_T^2 (p_1\!\cdot\!p_2)} 
  +
S_{8}^{\mu\nu\rho\sigma}  \frac{1}{p_T^2} 
  &&\\\nonumber  & && \;-&&
S_{9}^{\mu\nu\rho\sigma} \frac{(p_1\!\cdot\!p_3) (p_2\!\cdot\!p_3)}{p_T^2 (p_1\!\cdot\!p_2)^2} 
  -
S_{10}^{\mu\nu\rho\sigma} \frac{p_T^2 (p_1\!\cdot\!p_2)+(p_2\!\cdot\!p_3)^2}{p_T^2 (p_1\!\cdot\!p_2)^2} 
  -
S_{11}^{\mu\nu\rho\sigma} \frac{p_T^2 (p_1\!\cdot\!p_2)+(p_1\!\cdot\!p_3)^2}{p_T^2 (p_1\!\cdot\!p_2)^2} 
  &&\\\nonumber  & && \;-&&
S_{12}^{\mu\nu\rho\sigma} \frac{(p_1\!\cdot\!p_3) (p_2\!\cdot\!p_3)}{p_T^2 (p_1\!\cdot\!p_2)^2} 
  -
S_{15}^{\mu\nu\rho\sigma} \frac{(p_1\!\cdot\!p_3)}{p_T^2 (p_1\!\cdot\!p_2)} 
  -
S_{16}^{\mu\nu\rho\sigma} \frac{(p_2\!\cdot\!p_3)}{p_T^2 (p_1\!\cdot\!p_2)} 
  -
S_{18}^{\mu\nu\rho\sigma} \frac{1}{p_T^2 (p_1\!\cdot\!p_2)} 
  &&\\\nonumber  & && \;-&&
S_{19}^{\mu\nu\rho\sigma} \frac{1}{p_T^2 (p_1\!\cdot\!p_2)} \:.
\end{alignat}
\end{footnotesize}
These relations, as well as the inverse relations, are available in a computer-readable
format in Ref.~\cite{progdata}.

%- }}}

\FloatBarrier
\newpage

%- {{{ LO results for form factors:

\section{\label{app::T_LO}LO results for form factors}

\begin{figure}[h!]
    \includegraphics[width=0.98\textwidth]{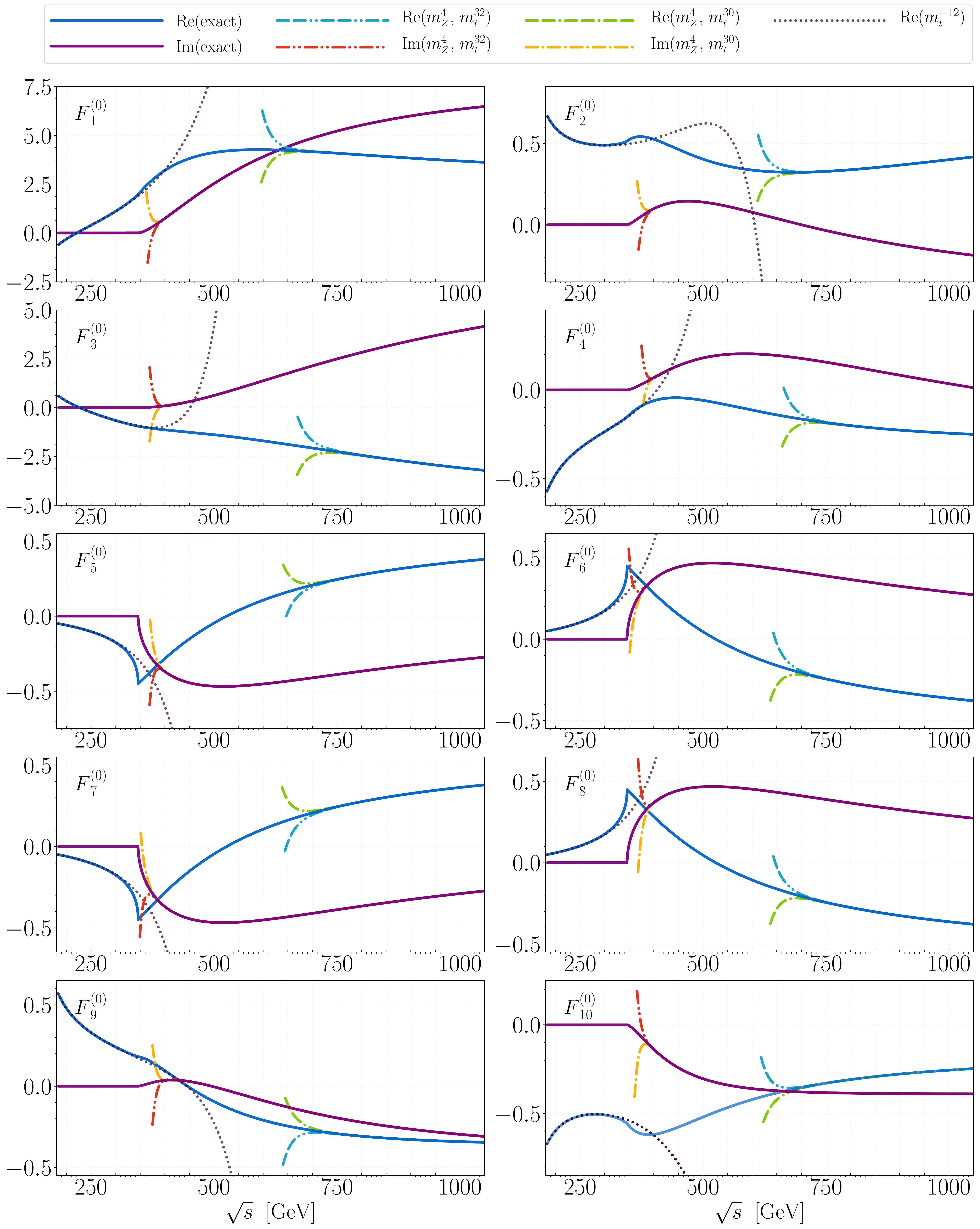}
  \caption{\label{fig::FF_LO_1}$F_1^{(0)},\ldots,F_{10}^{(0)}$
    as a function of $\sqrt{s}$, for $\theta=\pi/2$.}
\end{figure}

\begin{figure}[t]
    \includegraphics[width=0.98\textwidth]{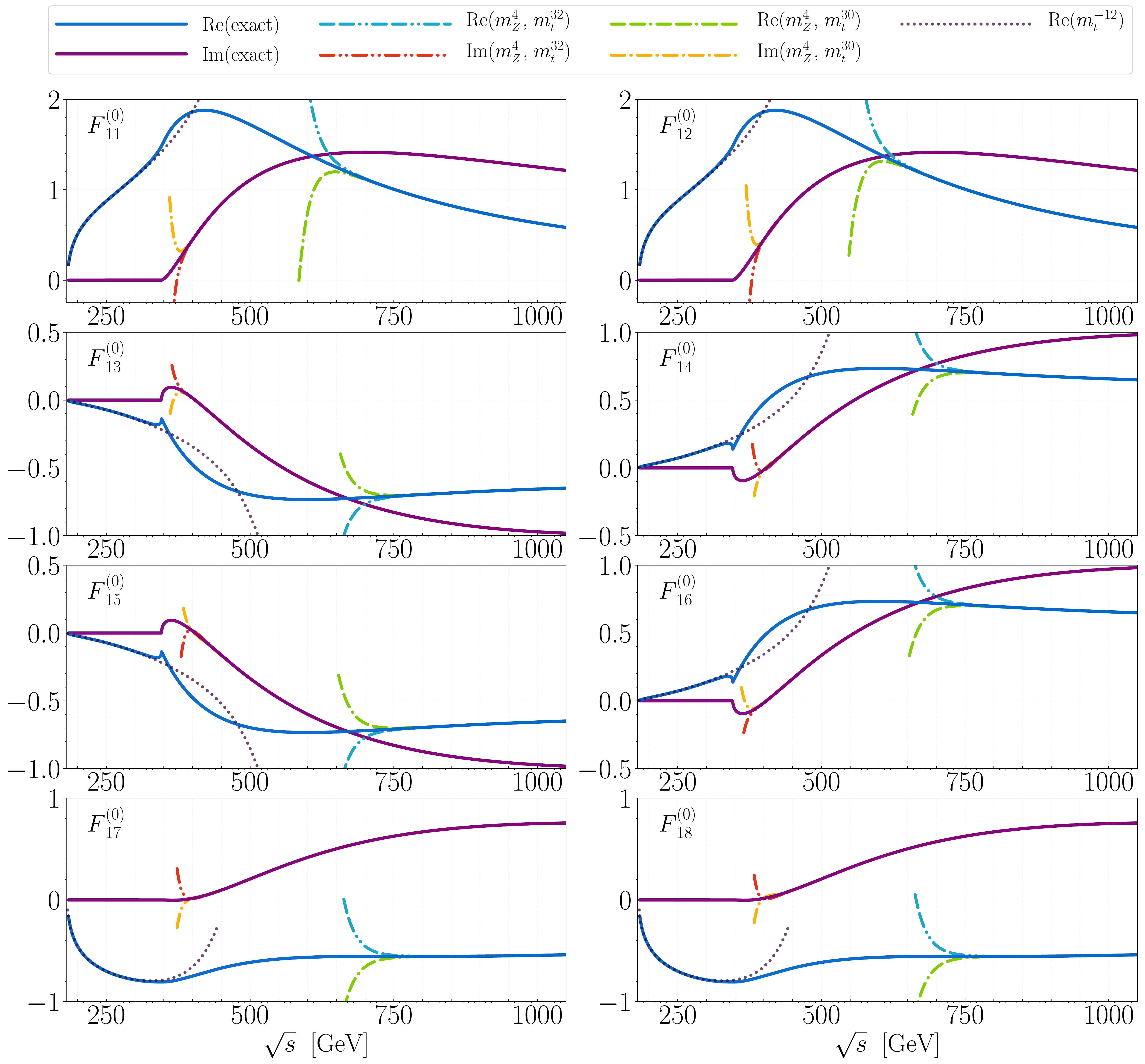}
  \caption{\label{fig::FF_LO_2}$F_{11}^{(0)},\ldots,F_{18}^{(0)}$
    as a function of $\sqrt{s}$, for $\theta=\pi/2$.}
\end{figure}

%- }}}

\FloatBarrier
\clearpage

%- {{{ NLO results for form factors:

\section{\label{app::T_NLO}NLO results for form factors}

\begin{figure}[h!]
    \includegraphics[width=0.97\textwidth]{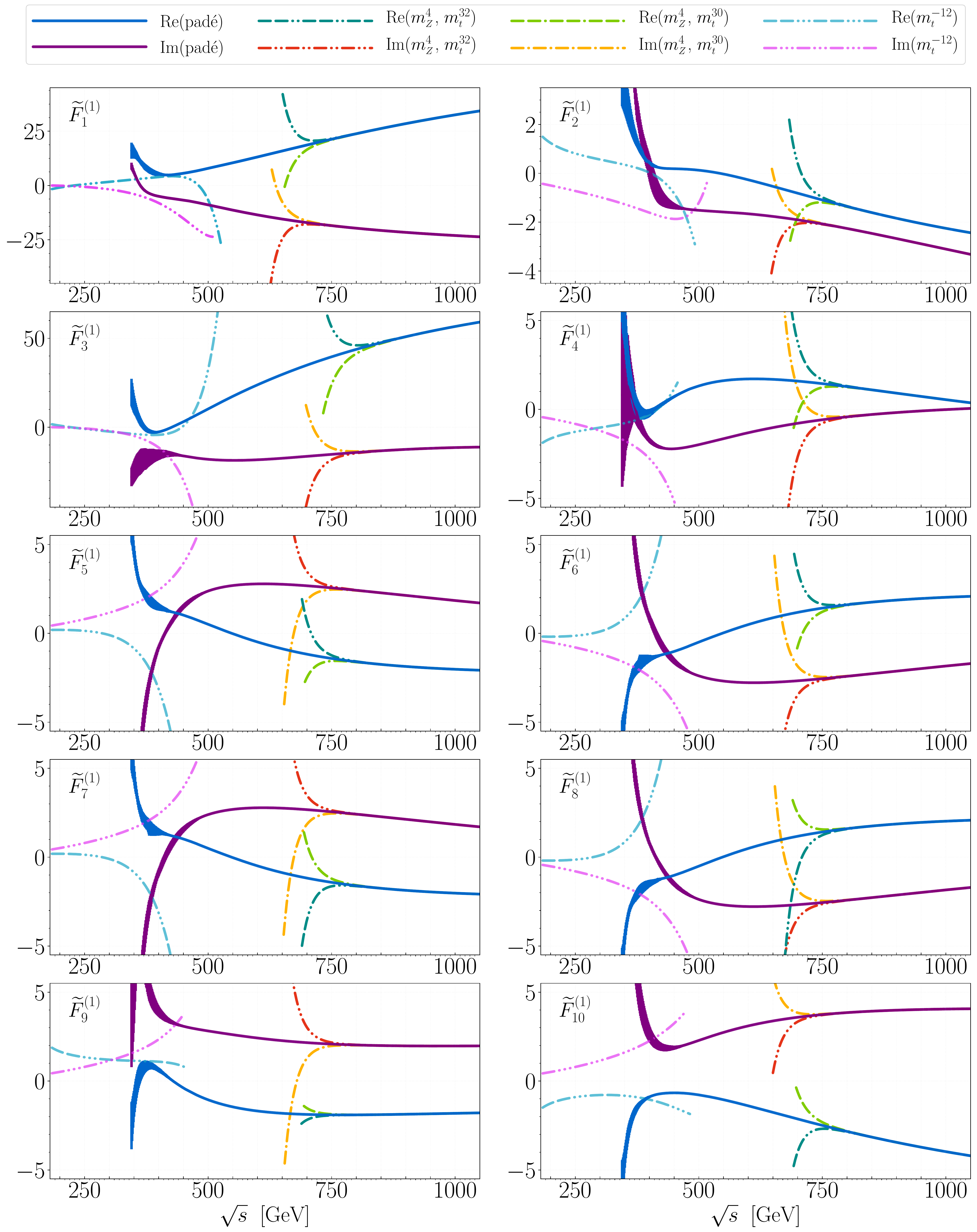}
  \caption{\label{fig::FF_NLO_1}$\widetilde{F}_1^{(1)},\ldots,\widetilde{F}_{10}^{(1)}$
    as a function of $\sqrt{s}$, for $\theta=\pi/2$.}
\end{figure}

\begin{figure}[t]
    \includegraphics[width=0.98\textwidth]{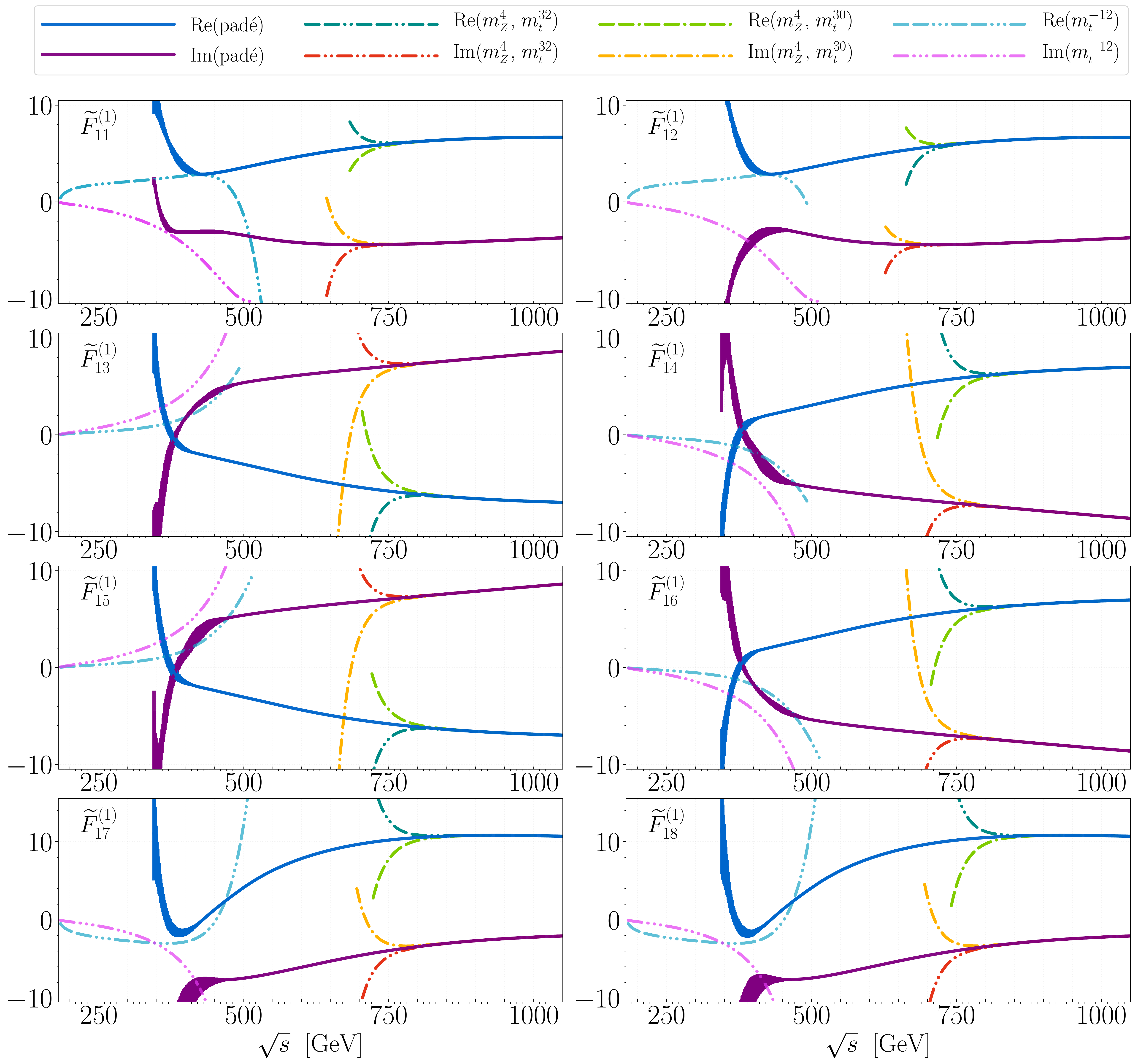}
  \caption{\label{fig::FF_NLO_2}$\widetilde{F}_{11}^{(1)},\ldots,\widetilde{F}_{18}^{(1)}$
    as a function of $\sqrt{s}$, for $\theta=\pi/2$.}
\end{figure}

%- }}}

\FloatBarrier

%- {{{ Analytic results:

\section{\label{app::example_analytic}Example analytic results for LO expansions}

In the ancillary file, we provide analytic expressions for the form factors defined in
Eq.~(\ref{eq::F_tri_box}), as expansions in both the large-$m_t$ and high-energy limits.
For the convenience of the reader we show here, for LO $f_1$ and $f_3$, the leading terms
of the expansions in a typeset form. In the large-$m_t$ limit they are given by
\begin{align}
f_{1,\mathrm{tri}}^{(0)}&=
-1
+\mathcal{O}\left(\frac{1}{m_t^{2}}\right)\:,\nonumber\\
f_{1,\mathrm{box}}^{(0),v_t}&= {v_t^2}
\left[\frac{3 s^2-14 s t-14 t^2}{360 m_t^4}+\mathcal{O}(m_Z^2)\right]+\mathcal{O}\left(\frac{1}{m_t^{6}}\right)\:,\nonumber\\
f_{1,\mathrm{box}}^{(0),a_t}&= {a_t^2}
\left[\frac{s}{3 m_t^2}+\mathcal{O}(m_Z^2)\right]+\mathcal{O}\left(\frac{1}{m_t^{4}}\right)\:,\nonumber\\
f_{3,\mathrm{box}}^{(0),v_t}&= {v_t^2}
\left[\frac{3 s^2+20 s t+3 t^2}{360 m_t^4}+\mathcal{O}(m_Z^2)\right]+\mathcal{O}\left(\frac{1}{m_t^{6}}\right)\:,\nonumber\\
f_{3,\mathrm{box}}^{(0),a_t}&= {a_t^2}
\left[-\frac{s}{6 m_t^2}+\mathcal{O}(m_Z^2)\right]+\mathcal{O}\left(\frac{1}{m_t^{4}}\right)\,.
\end{align}
and in the high-energy limit by
\begin{align}
f_{1,\mathrm{tri}}^{(0)}&=
m_t^2 \left(\frac{3 l_{ms}^2}{2 s}-\frac{6}{s}\right)+\mathcal{O}\left(m_t^{4}\right)\:,\nonumber\\
f_{1,\mathrm{box}}^{(0),v_t}&={v_t^2}
\left[
\frac{l_{1ts}^2 (s+t) \left(s^3+3 s^2 t+t^3\right)}{6 s^2 t^2}
+\frac{l_{ts} t^2}{3 s^2+3 s t}-\frac{l_{ts} l_{1ts} \left(3 s^2+2 s t+2 t^2\right)}{6 s^2}
\right.
\nonumber\\
&
-\frac{1}{3}+\frac{\pi ^2 \left(2 s^6+12 s^5 t+21 s^4 t^2+20 s^3 t^3+15 s^2 t^4+6 s t^5+2 t^6\right)}{12s^2 t^2 (s+t)^2}
\nonumber\\
&
\left.
-\frac{l_{1ts} (s+t)^2}{3 s t}
+\frac{l_{ts}^2 t \left(3 s^3+6 s^2 t+3 s t^2+t^3\right)}{6 s^2 (s+t)^2}
+\mathcal{O}\left(m_Z^{2}\right)
\right]
+\mathcal{O}\left(m_t^{2}\right)\:,\nonumber\\
f_{3,\mathrm{box}}^{(0),v_t}&={v_t^2}
\left[
\frac{l_{1ts}^2 \left(s^4-2 s^3 t-3 s^2 t^2-2 s t^3+t^4\right)}{6 s^2 t^2}
+\frac{l_{ts} l_{1ts} \left(3 s^2+4 s t-2 t^2\right)}{6 s^2}
\right.
\nonumber\\
&
+\frac{l_{ts} t^2}{3 s^2+3 s t}
+\frac{1}{3} l_{1ts} \left(-\frac{s}{t}-\frac{t}{s}+1\right)+\frac{l_{ts}^2 t \left(-3 s^3-3 s^2 t+t^3\right)}{6 s^2 (s+t)^2}
\nonumber\\
&
\left.
-\frac{1}{3}+\frac{\pi ^2 \left(2 s^6-9 s^4 t^2-16 s^3 t^3-9 s^2 t^4+2 t^6\right)}{12s^2 t^2 (s+t)^2}
+\mathcal{O}\left(m_Z^{2}\right)
\right]
+\mathcal{O}\left(m_t^{2}\right)\,,
\end{align}
where $l_{ms}=\log\left(m_t^2/s\right)+i\pi$, $l_{ts}=\log\left(-t/s\right)+i\pi$ and
$l_{1ts}=\log\left(1+t/s\right)+i\pi$.
Here we show only $f_{1,\mathrm{box}}^{(0),v_t}$ and $f_{3,\mathrm{box}}^{(0),v_t}$ since,
for the leading terms, they are equal to $f_{1,\mathrm{box}}^{(0),a_t}$ and
$f_{3,\mathrm{box}}^{(0),a_t}$ up to the replacement
$v_t^2 \to a_t^2$ as explained below Eq.~(\ref{eq::F_tri_box}).

%- }}}

\end{appendix}

%- }}}

\end{document}